# Effectiveness of Testing, Tracing, Social Distancing and Hygiene in Tackling Covid-19 in India: A System Dynamics Model

Jayendran Venkateswaran[1] and Om Damani[2]
[1] Industrial Engineering and Operations Research,
[2] Department of Computer Science, and associated with Center for Policy Studies,
IIT Bombay

18th April 2020

---

## Abstract

We present a System Dynamics (SD) model of the Covid-19 pandemic spread in India. The detailed age-structured compartment-based model endogenously captures various disease transmission pathways, expanding significantly from the standard SEIR model. The model is customized for India by using the appropriate population pyramid, contact rate matrices, external arrivals (as per actual data), and a few other calibrated fractions based on the reported cases of Covid-19 in India. Also, we have explicitly modeled, using independent time-variant levers, the effects of testing, contact tracing, isolating Covid-positive patients, quarantining, use of mask/better hygiene practices, social distancing through contact rate reductions at distinct zones of home(H), work(W), school(S) and other(O) locations. Simulation results show that, even after an extended lock-down, some non-trivial number of infections (even asymptomatic) will be left and the pandemic will resurface. Only tools that work against the pandemic is high rate of testing of those who show Covid-19 like symptoms, isolating them if they are positive and contact tracing all contacts of positive patients and quarantining them, in combination with use of face masks and personal hygiene. A wide range of combination of effectiveness of contact tracing, isolation, quarantining and personal hygiene measures help minimize the pandemic impact and some imperfections in implementation of one measure can be compensated by better implementation of other measures.

## Acknowledgements

*We acknowledge the support and contributions of Pooja Prasad, Shreenivas Kunte, Vanessa Beddoe, Rishav Deval and Priyesh Gupta in data collection from various sources and preliminary analysis.*

## 1. Introduction

A mathematical model of Covid-19 pandemic in India, based on System Dynamics (SD) methodology is presented. Our SD based epidemiological model explicitly captures the disease progression pathways, intervention pathways, causal linkages, and assumptions. Systems Dynamics approach has its origins in the study of industrial dynamics by Jay Forrester of MIT, in the late 1950s. In 1960s and 70s, several seminal works such as *Limits to Growth, Urban Dynamics,* and *World Dynamics* emerged. More recently, the *C-ROADS World Climate Policy model*, developed using an SD approach to understand the long-term climate effects, has been widely adopted.

SD Philosophy emphasizes capturing real world processes, causal pathways and endogenous feedback. The structure of an SD model must correspond to real world entities at appropriate level of abstraction and aggregation. The entity flows in the model must respect conservation laws, resource constraints, and known delays in real life. The parameters of the model must have a corresponding real-world interpretation and must come from domain knowledge, or, when estimated from data, must fall within plausible range.

Our model is an extension of the classic SEIR (Susceptible-Exposed-Infectious-Recovered) model (Anderson and May 1992). It is a compartmental model of the pandemic, wherein the population is divided into various age group wise 'compartments' to indicate Covid-19 disease progression and the interventions. Within each compartment, population is assumed to be homogeneous without and spatial or temporal attributes. While there are a number of SIR based models (Lourenco et. all. 2020; Ghaffarzadegan and Rahmandad 2020, MedRxiv 2020), our model differs from them not just in much more detailed modeling of various stages of disease progression, but also differs in explicit modeling of quarantining and isolation of both symptomatics and asymptomatics under key interventions of Testing and Contact Tracing. We also have a much more detailed model of the healthcare subsystem that is absent in almost all other models. Unlike (Ghaffarzadegan and Rahmandad 2020), we do not assume any seasonality effect of temperature and humidity since the evidence on it is thin and we would rather be conservative.

Given the time pressure, rapidly evolving data and uncertainties about virus infectivity and duration of various disease progression stages, SD is a good choice for evaluating policy interventions since it allows easy enlargement of model boundaries to incorporate hygiene practices, healthcare staff shortage, and behavioral feedbacks. The accuracy of absolute numbers generated by the model depend greatly on how accurate structural assumptions, assumed/estimated parameters, and input data, are. As long as the parameters and input data get the relative numbers and trends right, we can compare the relative impact of various policy interventions and learn which interventions or intervention combinations are most likely to take the system to desired state. This is particularly critical when almost simultaneously estimation of number of current infections in UK from very well-respected research teams (Lourenco et. all. 2020; Ferguson et. al. 2020) differ by orders of magnitude. Similar order of magnitude difference in number of infections are noted for Santa Clara county in Califorina (Bendavid, et al. 2020)

SD framework makes incorporating behavioral feedback possible. For example, as the number of infections cases rise, people may adopt better personal hygiene. Behavioral feedback is also important to model the counter-intuitive behavior of social systems and possible policy resistance. The success of a prolonged lockdown may come at the cost of wearing people down and they may stop following any containment measures resulting in a much worse second wave of pandemic. Also, the strict Covid-containment measures may inflict high collateral damage from other diseases. The devastating economic impact of strict measures may make people desperate and more vulnerable if their health and nutritional levels suffer. Due to time pressure many such scenarios are not incorporated here but must be kept in mind when making policy choices.

The rest of the report is organised as follows. First, we list out the assumptions as they also define the ability of our model. Next, we broadly describe our proposed model, along with the model validation/ usage of data. Third, we present our experimental findings based on simulations for multiple states of India. We then conclude with a brief discussion of our findings.

## 2. Modelling Assumptions

Like most models of SIR family, we also model the probability of movement of people from one compartment to other in a given time-step as independent of the residence time. Hence ours is a memory-less model in that residence time in each compartment is exponentially distributed. However, note that infectiousness of an individual is not uniform between exposure and recovery. It is initially negligible and gets hold up to 48 hours before the symptoms set in (Bai et. al. 2020; He and Lou et. al. 2020). As suggested in (Jacquez and Simon 2002), we take care of the limitations of the exponential distribution by introducing by breaking 'infectious' compartment in exposed, asymptomatic infectious, and Infectious symptomatics.

General assumptions with regards to our model are as follows:

- The population is assumed to be homogeneous and well-mixed. And there is no significant change on the total population due to births and deaths.
- The population (people) interacts at various locations that can be broadly divided into four sphere/ zone: at home, at work, at school and at other locations.
- The interactions are age-dependent. For example: we can expect high interaction among 5-20 year olds at schools, while 30-60 year olds may interact more at work.

- Patients infected by the coronavirus take some time (incubation period of ~ 5 days) before they show any symptoms. During this asymptomatic/ pre-symptomatic phase, they are assumed infective a few days prior (~24 to 48 hours prior) to the onset of symptoms.
- Symptomatic individuals may recover after a mild illness, or become serious and require hospitalisation. Serious patients may recover or can become critical (requiring ventilator / ICU facilities). The critical patients can recover or, unfortunately, die.
- The infectiveness of the asymptomatic individuals are assumed to be lower than that of the symptomatic individuals.
- Asymptomatic individuals may also recover from the disease without showing symptoms.
- The progression of the disease is age-dependent. Co-morbidity is assumed to be implicitly captured by this.
- The new cases are assumed to be detected only when the individuals are tested positive. This happens when either, (i) mild symptomatic individuals are tested positive, or (ii) the disease worsens and the patient seeks treatment at hospital, at which point it in time, they are tested and found positive.

Assumptions with regards to India-specific pandemic progress are as follows:

- The data about state-wise daily reported cases, recovered patients, and deaths, was obtained from www.covid19india.org.
- The disease is assumed to be *imported* to the country through arriving passengers only (as per the 'imported cases' reported in above link).
- It is assumed that the weather & climatic conditions of India play no role in the disease transmissions/ infectiousness. Also, it is assumed that the immunity level of Indians is similar to the people across the globe; and vaccinations such as BCG etc does not affect the infectiousness of the disease.
- Lockdown is assumed to reduce the interaction at work and other locations by 80% (with the remaining 20% representing essential services and front-line workers), and also result in 100% reduction in school zone interactions.
- The naïve ratio of the total number of deaths divided by the total cases reported, is at 3.4% for India which is very high. The case fatality ratio depends on the duration of illness and is hard to estimate without accurate patient records. The high deaths reported in India could be due to delay in seeking treatment by the patients (due to which the disease might have progressed to critical state thus hastening death), or due to the prior medical conditions of the patients, or due to delays in initial testing and reporting.

Other modelling assumptions:

- In national level 'All-India' model, the interaction among states is assumed to be implicit.
- The state level models are considered independent of each other, i.e., no interactions among states is assumed.

- Super spreader events are not explicitly considered.
- Economic effects are not directly considered.
- Once lockdown is lifted, no further imported cases are assumed to occur. This can be thought of as a policy choice. While some researchers have argued that once community transmission sets in, imported cases do not make much difference (Chinazzi et. al. 2020), recent experience of Vietnam, Singapore and China seem to contradict it. Hence we take no new import as a policy decision for the current scope of our work.

Definition Used:

- *Infectivity*: It represents the probability of an individual contracting the disease upon interaction with an infected person.
- *Contact Rate*: It represents the average number of persons a person interacts with in a day.

These two definitions are particularly important in that they relate to the concept of the popular but often misunderstood $R_0$, the basic Reproduction Number. It refers to the expected number of individuals, a single infected individual directly infects when introduced in a fully susceptible population, in absence of external interventions. Its average value is computed as the product of the number of interactions per day between an infected and a susceptible person, the probability of infection in such an interaction, and the duration where the person stays infectious. Human to fomite (inanimate infectious objects) contacts are usually not modelled separately and are supposed to be subsumed in these averages.

This tells us that to reduce disease transmission we can target any of the three components of $R_{0.}$ Interventions like social distancing, quarantine, isolation, and lockdowns affect the number of interactions, medical interventions can reduce the disease duration, while the use of masks, PPE, and better hygiene can reduce the infectivity. Infectivity is thought to be a property of the virus but it also depends on the intensity of the interactions (in addition to the virus property, and immunity against virus, and other conjectures such as temperature/humidity etc). Both the number of interaction and its intensity (as a function of physical distance) is affected by population density. Hence both the social customs and behaviours and spatial characters of a region will affect $R_0$. All of this means $R_0$ computed for a region cannot be blindly used in another region. In this work, we address these issues by calibrating the model for contact rates and impact of hygiene on infectivity.

## 3. Model Description

A high level stock-flow diagram of the proposed model is shown in Figure 1. We assume the population of the region is age-stratified into *M* age-groups. The population within each age group is divided into *Susceptibles (S)*, *Exposed(E)*, *Asymptomatic infectives (A)*, *Infectious Symptomatic (I)* cases, *Hospitalised patients (H)*, *Critical patients (C)*, *Recovered (R)* population and the *Dead (D).* The governing equations of the model are described in detail in Appendix A.

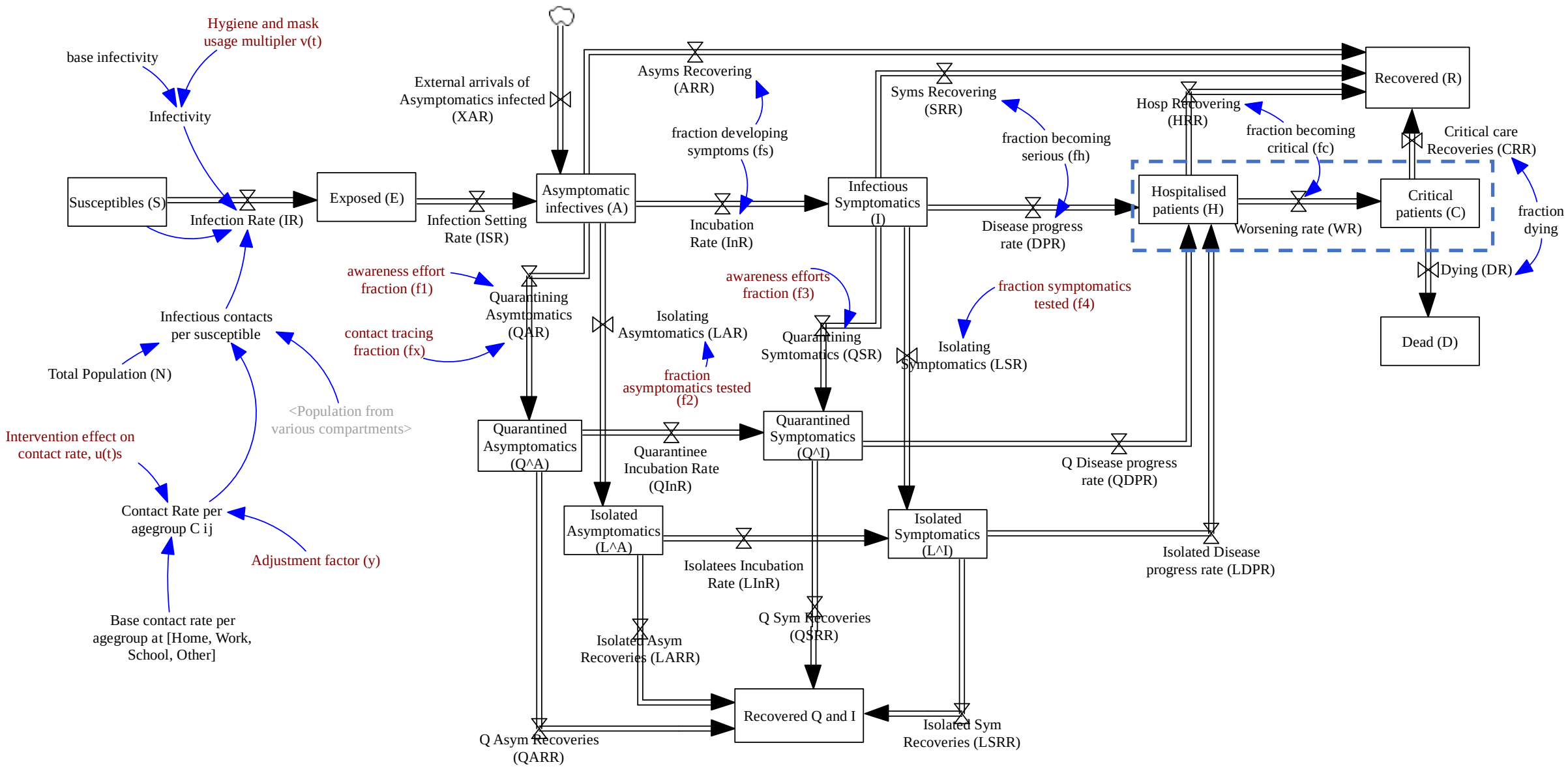

*Figure 1: High level stock flow diagram of the proposed model for Covid-19 India pandemic. The intervention points are represented by causal variables shown in red color. Other causal variables, especially delays, are not shown to improve readability.*

Let's consider the 'primary flow' of the diseases across the compartments, that is, ignoring quarantine and isolations. The *Susceptible* (S) people get exposed to the virus from contact with an infectious person. Initially all those exposed to virus are non-symptomatic and non-infectives, and become infective (but not symptomatic) only after a latency period of $d_{EA}$ days. After another $d_{AI}$ days, these individuals develop symptoms and move to *Infectious Symptomatics* compartment. The sum of $d_{EA}$ and $d_{AI}$ represents the average incubation period, $d_P$. The persons in *Infectious Symptomatics* compartment are said to exhibit mild symptoms and are active in the community, and thus continue to transmit the virus. A fraction of such mild cases recovers, while for others, the disease worsens resulting in hospitalization (move to *Hospitalised patients* class). A fraction of the serious hospitalised patients recovers, while the rest become critical requiring ventilator / ICU facilities (move to *Critical patients* class). A fraction of the critical patients recover while the rest, unfortunately, die. It is assumed that the *Recovered* patients and the *Dead* do not transmit the virus. Note that the external seeding of the virus in the community is captured explicitly.

Now, suppose quarantining and isolations of infectious persons are possible. This is explicitly captured by the flows out of the stocks A and I. The flow to the Quarantine and Isolation stocks are modeled as governed by different control fractions. The flow of asymptomatics from *A* to $Q^A$ represents the proportion of persons who are quarantined either through awareness efforts (represented by $f_1$(t)) or contact tracing of a positive tested patients (represented by $f_X$(t)). The flow of asymptomatics from *A* to $L^A$ represents the proportion of persons who are isolated after testing positive for Covid-19 though early tracing and testing effort (represented by $f_2$(t)). The quarantined/ isolated asymptomatics, either recovers, or develop symptoms and proceed to the quarantined Symptomatic and Isolated Symptomatic stocks. The flow of Infectious Symptomatics from *I* to $Q^I$ represents the proportion of persons who are quarantined through awareness efforts (represented by $f_3$(t))). The flow of symptomatics from *I* to $L^I$ represents the proportion of persons who are isolated after testing positive for Covid-19 (represented by $f_4$(t)). This, $f_4$(t), represents the fraction of symptomatic individuals who are identified during the initial stages of the symptoms itself. The quarantined/ isolated symptomatics, either recover, or become serious and are hospitalised. Note that the patients who are not quarantined or not isolated proceed along the primary flow a described earlier. The flow equations are given by equations 22-52 in Appendix A. Also, note that all the flows are governed by explicit delays between stages, some of which are determined by the nature of the disease (viz., incubation period), while others are determined by the intervention or control action (viz., testing delay).

The infection rate is governed by the contact rate (λ) and the infectivity (infection probability, *p*), as shown in Equations 16-21 in Appendix A. The contact rate for an age-group is the sum of all interactions of that age group with all other age groups across all contact locations (home, work, school and other). It is assumed that a susceptible individual can get infected upon contact with an *Asymptotic infective,* or *Infectious Symptomatics* person, including those in quarantine and isolation. It is assumed that the relative infectivity of the *Asymptomatic infectives* is only (1-*k*)% of the symptomatic infected class. Further, it assumed that patients in hospitals and critical care are in perfect isolation and hence do not contribute to the disease transmission.

Social distancing measures, such as school closures, lockdown, partial opening of work, etc, are modeled using fractions $u^H(t)$, $u^W(t)$, $u^S(t)$, $u^O(t)$, which represent the time dependent control measure imposed on contacts at home(H), work(W), school(S) and other(O) locations, respectively. Reductions in the infectivity due to measures such as personal hygiene, usage of PPE such as masks, etc., is modeled using fraction multiplier $v(t)$ on the base infectivity parameter *p*. Additionally, a time dependent adjustment factor *y*(t) is used as a multiplier of contact rate, $C_{ij}$. At the start of the Covid-19 pandemic there is typically little awareness about the diseases, and hence little or no precautions are taken by the population. This can result in an increased interaction, more than what would have otherwise been.

Further, in the model, the overflows that can possibly occur at hospitals is explicitly modelled. The area shown by the dashed box in Figure 1 is expanded in the main model, as shown in Figure 2. The overflows at hospitals and Critical care facilities are explicitly modelled to capture the possible deterioration in the care due to overwhelming demand on the hospital. The flows *DPR,* QDPR, LDPR, represent the 'new arrivals' to the hospitals, feeding into the stock DH. Now, as per the available capacity ($K^H$) at the hospital, that many patients are sent immediately to the hospital stock H, while the rest are explicitly marked as overflow hospital patients and moved to stock OH. (Note: both these class of patients could be physically at the hospital itself). Similarly, the patients at the hospital (stock H), as their condition worsens, can go to critical care facility if there is sufficient capacity ($K^C$). If not, the patient 'move' to overflow critical patients (OC). The OH patients can worsen (become critical) or recover, where the fraction recovering could be lower than that for the regular 'H' patients. The OC patients can die or recover, where the fraction recovering could be lower than that for the regular 'C' patients.

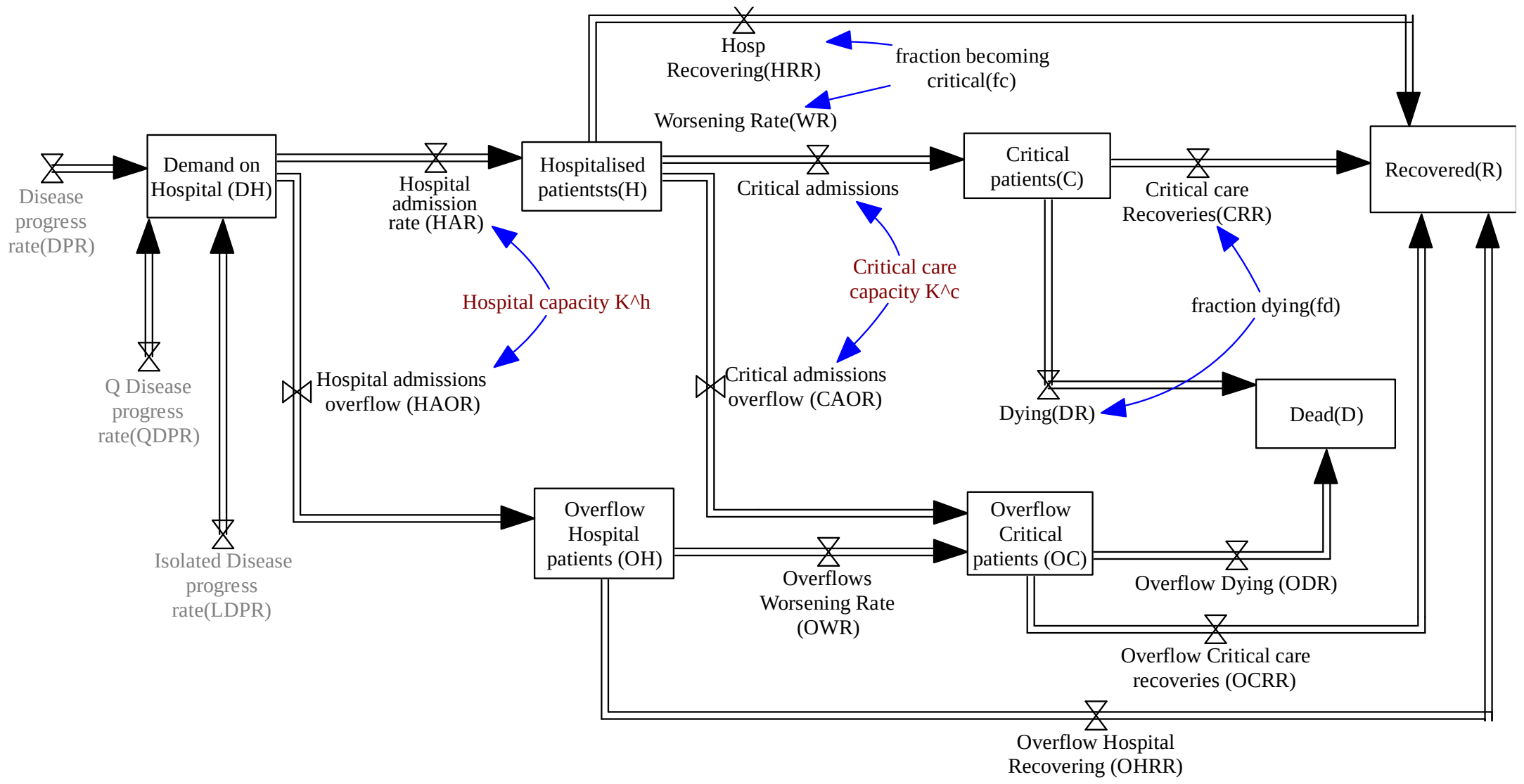

*Figure 2: Expanded view of the hospital incorporating overflows*

## 4. Parameter settings

The model parameter settings are based on published reports and web sources (see Assumptions section), as follows:

*Table 1: Parameter settings used*

| **Parameters** | **Value(s)** | **Source** |
|---|---|---|
| **Population parameter** | | |
| Age-groups, *i* | 16 (with bands of 5 years starting at 0. The last band is assumed to be include all who are 75+) | |
| Population fraction, $g_i$ | 0.086, 0.088, 0.093, 0.092, 0.089, 0.085, 0.084, 0.074, 0.064, 0.057, 0.049, 0.043, 0.035, 0.026, 0.017, 0.018; | Census 2011 |
| Populations | Population of all states, taken as 2018 values, from: http://statisticstimes.com/demographics/population-of-indian-states.php | |
| **Fractions** | | |
| Fraction developing symptoms, $f_i^S$ | 0.8, for all age groups | Verity et. al. 2020 |
| Fraction becoming serious, $f_i^H$ | 0.02, 0.02, 0.02, 0.02, 0.17, 0.17, 0.17, 0.17, 0.17, 0.28, 0.28, 0.27, 0.27, 0.4, 0.4, 0.47 | CDC report 2020 |
| Fraction becoming critical , $f_i^C$ | 0, 0, 0, 0, 0.14, 0.14, 0.14, 0.14, 0.14, 0.23, 0.23, 0.24, 0.24, 0.28, 0.28, 0.34.<br>(fraction $f_i^{CO}$= min(0.8, 2*$f_i^C$, ) | |
| Fraction dying, $f_i^D$ | 0.02, 0.02, 0.02, 0.02, 0.06, 0.06, 0.06, 0.06, 0.06, 0.11, 0.11, 0.25, 0.25, 0.26, 0.26, 0.3<br>(fraction $f_i^{DO}$= min(0.9, 2*$f_i^D$) | |
| **Infectivity and Contacts** | | |
| Base infectivity, $\hat{p}$ | 0.015 | Singh and Adhikari 2020 |
| Reduction in infectiousness of asymptomatics, k | 0.5 | Average of various sources |
| Contact Matrices | See Appendix B | |
| **Delay Parameters** | | |
| Incubation period, $d_{EA} + d_{AI}$ | 5 days | Personal communication with doctors; Multiple source |
| Latency period, $d_{EA}$ | 3 days | |
| Avg. time to develop symptoms, $d_{AI}$ | 5 – 3 = 2 days | |
| Avg. time for asymptomatics to recover, $d_{AR}$ | 11 days | |
| Avg. time for infectious Symptomatics to become serious/ hospitalise, $d_{IH}$ | 5 days | |
| Avg. time for infectious Symptomatics to recover, $d_{IR}$ | 14 days | |
| Avg. worsening duration, $d_{HC}$ | 5 days | |
| Avg. time for hospitalised patients to recover, $d_{HR}$ | 14 days | |
| Avg. time for critical patients to die $d_{CR}$ | 5 days | |
| Avg. time for critical patients to recover $d_{CR}$ | 14 days | |
| Avg. time to quarantine / isolate asymptomatics, $d_{AQ}, d_{AL}$ | 1 day (assumed half of $d_{AI}$) | |
| Avg. time to quarantine / isolate symptomatics, $d_{IQ}, d_{IL}$ | 2.5 days (assumed half of $d_{AI}$) | |
| Avg. time for quarantine / isolate asymptomatics to develop symptoms, $d_{QQ}, d_{LL}$ | 1 day<br>$(d_{AI} - d_{AQ})$ | |
| Avg. time for quarantine / isolate symptomatics to become serious/ hospitalise, $d_{QH}, d_{LH}$ | 2.5 days<br>$(d_{IH} - d_{IQ})$ | |
| Avg. time for quarantine / isolate asymptomatics to recover, $d_{QAR}, d_{LAR}$ | 10 days<br>$(d_{AR} - d_{AQ})$ | |
| Avg. time for quarantine / isolate symptomatics to become serious/ hospitalise, $d_{QIR}, d_{LIR}$ | 13 days<br>$(d_{IR} - d_{IQ})$ | |

## 5. Simulation and Model Validation

The model has been implemented using Vensim® Pro software. Some additional variables have been included in the Vensim model, mainly to capture a few of the underlying equations in parts. A simulation timestep of 0.0625 is used, with integration method of Euler. Results are saved per day.

The model has been thoroughly debugged to avoid integration time step errors, and floating point overflow caused by extreme small value (e.g. 0 in the denominator). The units have been checked to ensure consistency. Variable names have been chosen so that the model is understandable.

As part of model validation, direct structure tests have been conducted to ensure that relations & assumptions in the model are based on accepted theories and all important variables are included in the model. As per that, an essential feature included in our model is that a coronavirus infected person undergoes an incubation period before symptoms show up; and during the incubation period they are infectious (without showing symptoms) after a latency period. Other aspects included, which goes beyond the standard SEIR representation of Covid-19 are, the age-groups specific contacts across various contact zones or locations; age specific progression of & recovery from the disease; and explicit representation of quarantines and isolations. Behavior tests were conducted to ensure that the model behaves as per accepted System Dynamics theory at various test conditions, extreme values and loop knockouts.

## 6. Modelling the observed pandemic behaviour in India and Sensitivity Analysis

A conscious decision was taken by the modelers to <u>not</u> fine-tune the parameter values to exactly replicate the reported cases of Covid-19 in India, or to minimise some statistical measure of error. Instead it was decided to replicate the trend in behaviour by estimating a few parameters only (McCloskey and Ziliak 1996).

On the website www.covid19india.org, the India state-wise data on daily number of (newly) reported cases based on test reports, daily number of recovered cases, and daily number of deaths are reported. The first 3 recorded cases of Covid-19 infections were during Jan 30$^{th}$ – Feb 3$^{rd}$ 2020, followed by a long hiatus. Since March 2$^{nd}$ 2020, India has been seeing new cases reported daily. For modelling purposes the early February data is ignored. Also, the current deaths reported in India is quite high vis-à-vis the reported cases, and compared to the fatality rates reported in [CDC report]. This could be due to initial long delay in seeking proper treatment, comorbidity conditions, etc. Hence it is decided to calibrate the model based on the new cases reported only.

In our model, there are two pathways by which a new case of Covid-19 is identified/ confirmed via testing. One pathway is when the individual allows the disease to progress, and seek medical help only when it becomes severe. This is captured by the flow Disease Progress Rate (DPR) from Infectious Symptomatic (I) to Hospitalized Patients (H) compartments. Such patients are tested positive late after the onset of the disease. The second pathway will be to, via contact tracing, test some symptomatic individuals *a priori* before the symptoms become severe. This is captured by the flow Isolating Symptomatics Rate (LIR) from Infectious Symptomatic (I) to Isolated Symptomatics ($L^I$) compartments. This is governed by the fraction $f_4(t)$. Now, the data available in the website does not distinguish between the two pathways. It is important to do so, since pathway 1 is reactive, while pathway 2 is a proactive approach. It is assumed that pathway 1 is active be default, and hence only $f_4(t)$ is estimated in our model.

As far as the authors are aware, there are no detailed published clinical studies of the cases in India. In the absence of that, the parameters reported in Section 4.0 are assumed to hold true for India. Also, India has already started various interventions to mitigate/ prevent the pandemic spread. Based on reported studies and preliminary experiments, we decided estimate the following parameters only to replicate the trend in the daily reported cases:

(i) Control measures of $u^H(t)$, $u^W(t)$, $u^S(t)$, $u^O(t)$, to indicate the effect of lockdown. Prior to lockdown they were set at 1. During lockdown, assumed that $u^H(t) = 1$ and $u^S(t)=0$.
(ii) The proportion $f_4(t)$ of symptomatic infectious individuals who are isolated after testing.
(iii) The fraction $x(t)$ of asymptomatics traced and quarantined per fraction symptomatic tested. Note that fraction of asymptomatics traced and quarantined is given $f_x(t) = x(t)*f_4(t)$.
(iv) The time dependent adjustment factor $y(t)$.

The parameters, $f_1(t)$, and $f_3(t)$, are assumed to be 0, implying that volunteer quarantining through awareness doesn't happen. Also, $f_2(t)$ is assumed to be 0, implying a no testing policy of asymptomatic. Also, the time dependent control measure $v(t)$ for better hygiene and PPE use is kept at 1 to indicate no effect of this on Covid-19 spread in India.

<u>Introduction of infected individuals in the community</u>: Since the pandemic did not have origins in India. The spread in India is due to arrivals of infected individuals from abroad. Now, in the www.covid19india.org website (crowdsourced database available there), a few of the cases has been tagged as 'imported'. This is interpreted as those individuals who have arrived from abroad infected. Since we do not have the information on the exact arrival date to India, we assume that all these 'imported cases' came to India on an average of 7 $(d_{AI} + d_{IH})$ days prior to being identified as positive. Thus, for example, the imported patients on March 2nd is assumed to have arrived on February 24th. These imported cases arrivals is taken to define the flow $XAR(t)$. Also, lockdown is taken to start on 25th March 2020.

The results of model calibration exercise is shown in Figure 3 below, for the All India data. The simulation duration was from 24th Feb to 14th April. It is observed that the new cases and cumulative cases give a reasonable fit to the reported data of the pandemic's spread in India. Also, observe that the simulation over estimates the number recovered and underestimate the number dead. Further calibration of recovered and dead based on current data only has a grave risk of over-estimating deaths in future since it appears that people are approaching hospitals very late when disease has progressed significantly. We assume that this scenario will change as time passes. Also, in future we plan to take comorbidity and patient case details for Indian conditions into account.

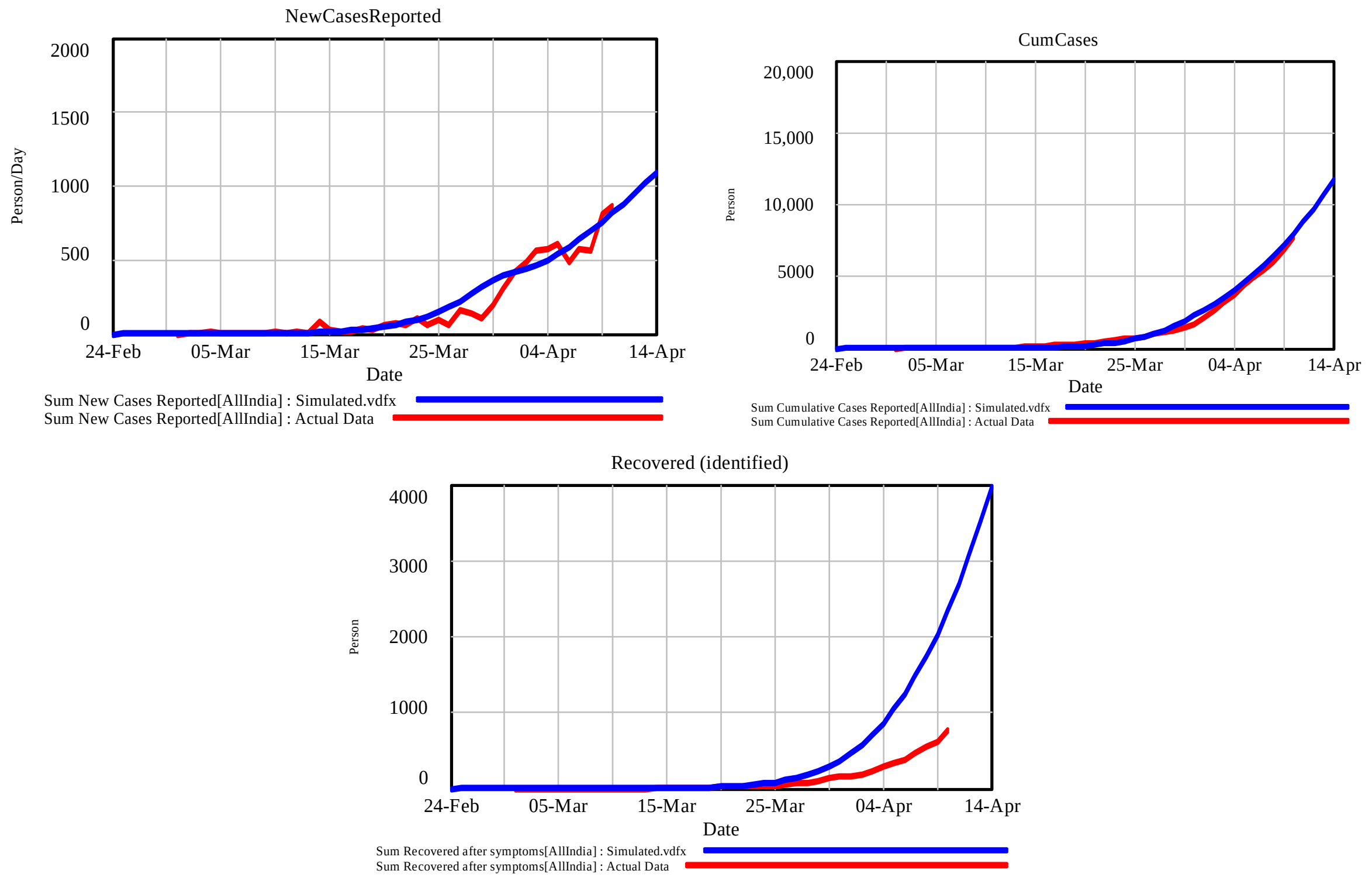

*Figure 3*: The comparison of simulated (blue line) and the actual data (red line), for the daily reported cases (top left), cumulative cases reported (top right), cumulative recovered (bottom).

The parameter setting that resulted in the above behaviour are:

(i) During lockdown, $u^H(t) = 1$, $u^W(t)$=0.3, $u^S(t)$=0, $u^O(t)$ =0.3, and 1 otherwise.

(ii) The proportion $f_4(t) = 0.25$

(iii) The proportion $x(t) = 0.1$

(iv) $$y(t) = \begin{cases} 2 & t \leq 15 \\ 2 - \left(\frac{0.8}{35}(t - 15)\right) & 15 < t < 50 \end{cases}$$

Next, sensitivity analysis of the model behaviour to the control parameters were performed, the results of which are shown in Figure 4 below. The sensitivity settings are as follows:

(i) $u^W(t)$ ~ UNIFORM(0.2, 0.4)

(ii) $u^O(t)$ ~ UNIFORM(0.2, 0.4)

(iii) $f_4(t)$ ~ UNIFORM(0.05, 0.5)

(iv) The slope of $y(t)$ ~ UNIFORM(0.8/35, 1/35)

It is observed that the model reasonably captures the reported trend even for changes in the calibrated values. During the initial days, the model seems to overestimate the trend. However, the more recent behaviour, especially post lockdown, is capture reasonably well. Hence, it is decided to calibrate the model with the recommended parameters.

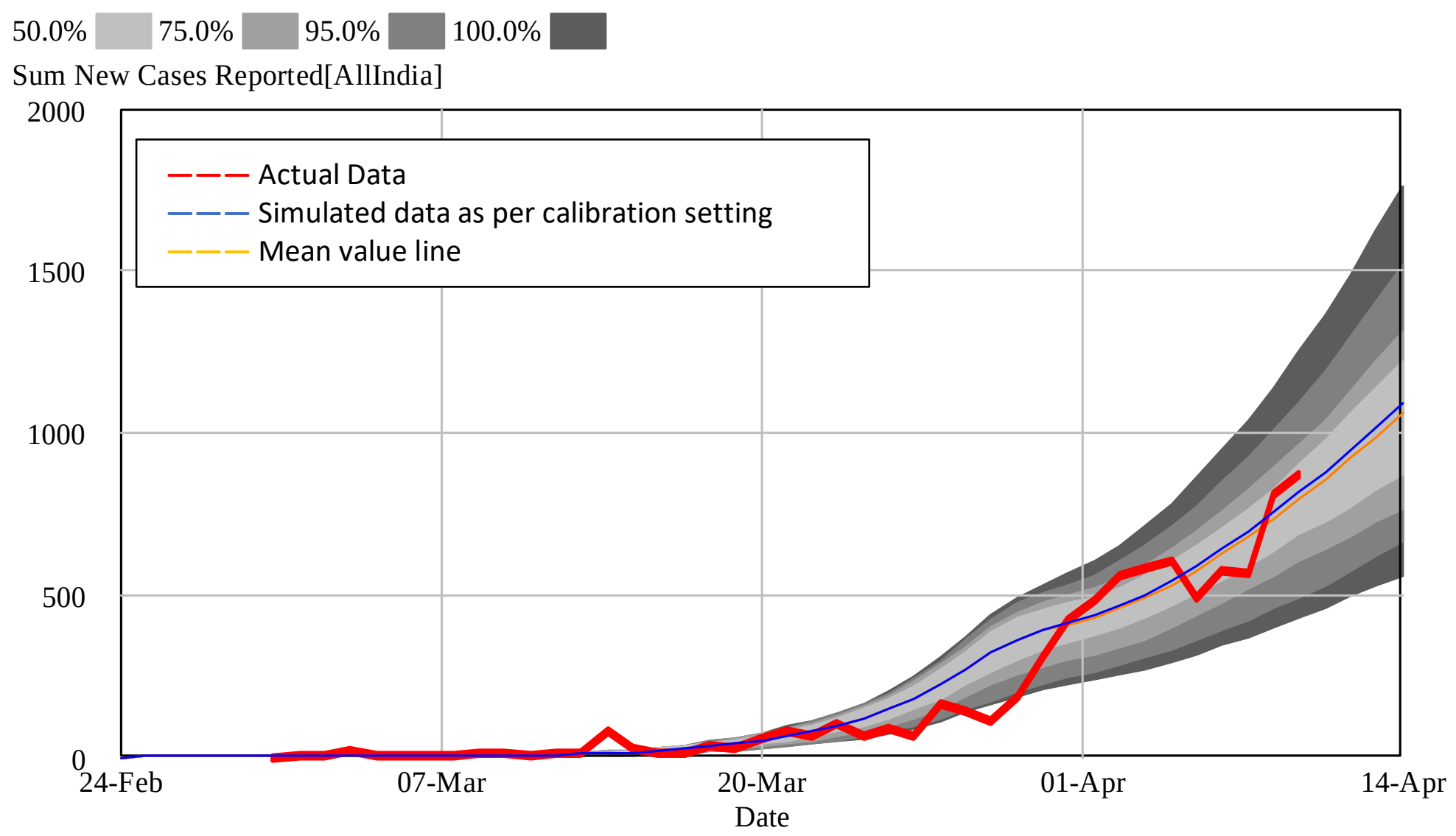

*Figure 4*: Sensitivity analysis confidence bounds (shown by the grey-scale bands) of New Cases Reported (India)

In a manner similar to what was describe for the all India data, the model was calibrated for each state (thus each state has a separate model). The results of state-wise calibration are shown in Appendix C.

## 7. Modelling the Interventions

This section describes the parameter settings used to model different intervention scenarios. All interventions are assumed to be effective with effect from April 15th. Until April 15th , the parameter setting obtained as in Section 6 are used.

*Table 2: Interventions considered*

| # | Intervention | Modeling, t>= Apr 15th |
|---|---|---|
| 1 | No extension of lockdown after 14th April | $u^H(t) = u^W(t) = u^S(t) = u^O(t) = 1$ |
| 2 | 50% reduction in contacts everywhere till 15th May | $u^H(t) = u^W(t) = u^S(t) = u^O(t) = 0.5$, from 15th Apr to 15th May; and 1.0 after that. |
| 3 | Extending the lockdown till 30th April | $u^H(t)=1$, $u^W(t)=0.3$, $u^S(t)=0$, $u^O(t)=0.3$, from 15th to 30th Apr; and 1.0 after that. |
| 4 | Extending the lockdown till 30th April and during 1st May to 15th May 50% reduction in contacts (same as 2) | $u^H(t)=0.5$, $u^W(t)=0.3$, $u^S(t)=0$, $u^O(t)=0.3$, from 15th to 30th Apr; $u^H(t) = u^W(t) = u^S(t) = u^O(t) = 0.5$, from 1st to 15th May; and 1.0 after that. |
| 5 | Extending lockdown till 15th May | $u^H(t)=1$, $u^W(t)=0.3$, $u^S(t)=0$, $u^O(t)=0.3$, from 15th Apr to 15th May; and 1.0 after that. |

| # | Intervention | Modeling, t>= Apr 15th |
|---|---|---|
| *Additional scenarios* | | |
| 6 | 50% reduction in contacts, & school closed | $u^H(t) = u^W(t) = u^O(t) = 0.5$, and $u^S(t)=0$ |
| 7 | 'Normal' contacts, and school closed | $u^H(t) = u^W(t) = u^O(t) = 1$, and $u^S(t)=0$ |

Each of the above interventions are simulated at two settings of testing and tracing:

A. With increased testing (testing and tracing; 80% symptomatic can be identified and isolated within one day of development of symptoms
   - Modelled with $f_4(t) = 0.8$, $x(t) = 1$, for t >= 15th Apr
   - Testing delay $d_{IL}$ continues to be 2.5 days.
   - $v(t)=0.8$, for t >= 15th Apr (minimal improved level of hygiene and masks usage)

B. Continue the same rate of testing
   - Modelled with $f_4(t) = 0.25$, $x(t) = 0.1$ for All India model. For all other states, the obtained calibrated values are used.
   - $v(t)=0.8$, for t >= 15th Apr (minimal improved level of hygiene and masks usage)

For the additional scenarios (policy 6 and 7) we have considered the following C and D settings also.

C. Along with (A), improved use of Face Mask and Personal Hygiene
   - Modelled with $f_4(t) = 0.8$, $x(t) = 1$, for t >= 15th Apr
   - Testing delay $d_{IL}$ continues to be 2.5 days.
   - $v(t) = 0.6$, for t >= 15th Apr

D. Modified (A), improved use of Face Mask and Personal Hygiene, and lower testing/tracing
   - Modelled with $f_4(t) = 0.5$, $x(t) = 1$ for All India model. For all other states, the obtained calibrated values are used.
   - Testing delay $d_{IL}$ continues to be 2.5 days.
   - $v(t) = 0.6$, for t >= 15th Apr

Thus, a total of 18 intervention scenarios are considered (1A, 1B, 2A, …, 5B, 6A, 6B, 6C, 6D, 7A, 7B, 7C, 7D), for the all India model and for each of the States.

## 8. Simulation Results and Observations

The above interventions were encoded, and the model simulated for India as well as each of the states. All simulations were run for a time period until June 30th 2020. Do note that it is assumed that no new cases arrive at the state/ country from an external source in the above simulation time period.

Figure 5 and 6 shows the simulation outputs for all India data for intervention scenarios 1A to 5B. Figure 5 shows that interventions 1B, 2B, 3B, 4B, and 5B (4B and 5B overlaps in figure) are significantly more than their 'A' counterparts (all overlapping, along the x-axis). This clearly shows that increased testing and tracing and isolating has a significant impact in controlling the pandemic spread.  Let's zoom in (see Figure 6). It can be readily observed from Figure 6 that once lockdown is lifted (any of the scenarios 1-5), the new cases reported will increase exponentially (all 'B' scenarios). The policies 2A, 4A and 5A  seem to result in low spread of pandemic, with policy 4A being the least.  Also, it is observed that all the 'A' policies, 1A to 5A will result in a short term increase in the number of reported cases. This is due to the larger number of cases that gets discovered initially due to increase in the testing rates.

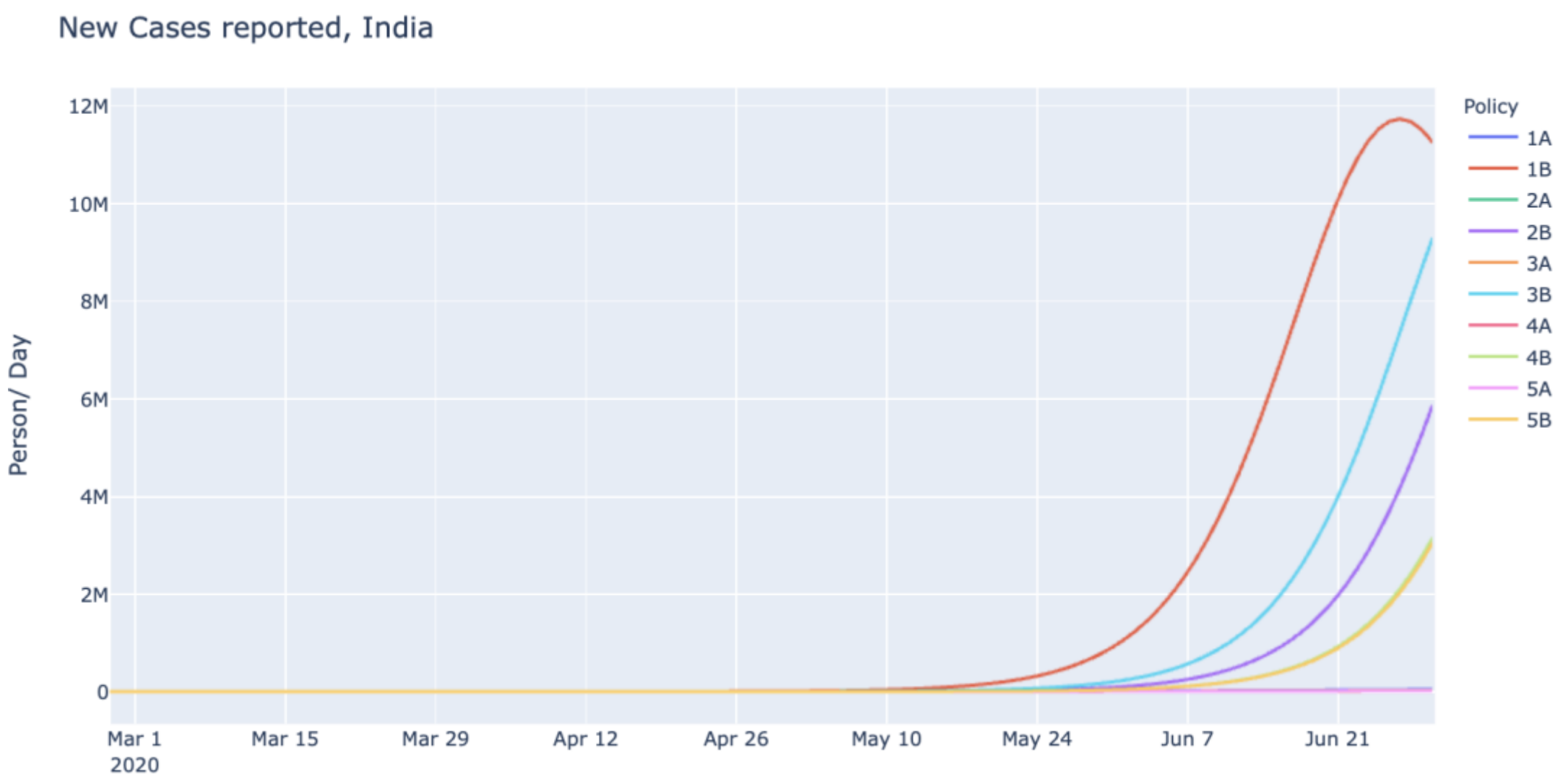

*Figure 5: Simulation response based on India data (until June 30th 2020).*

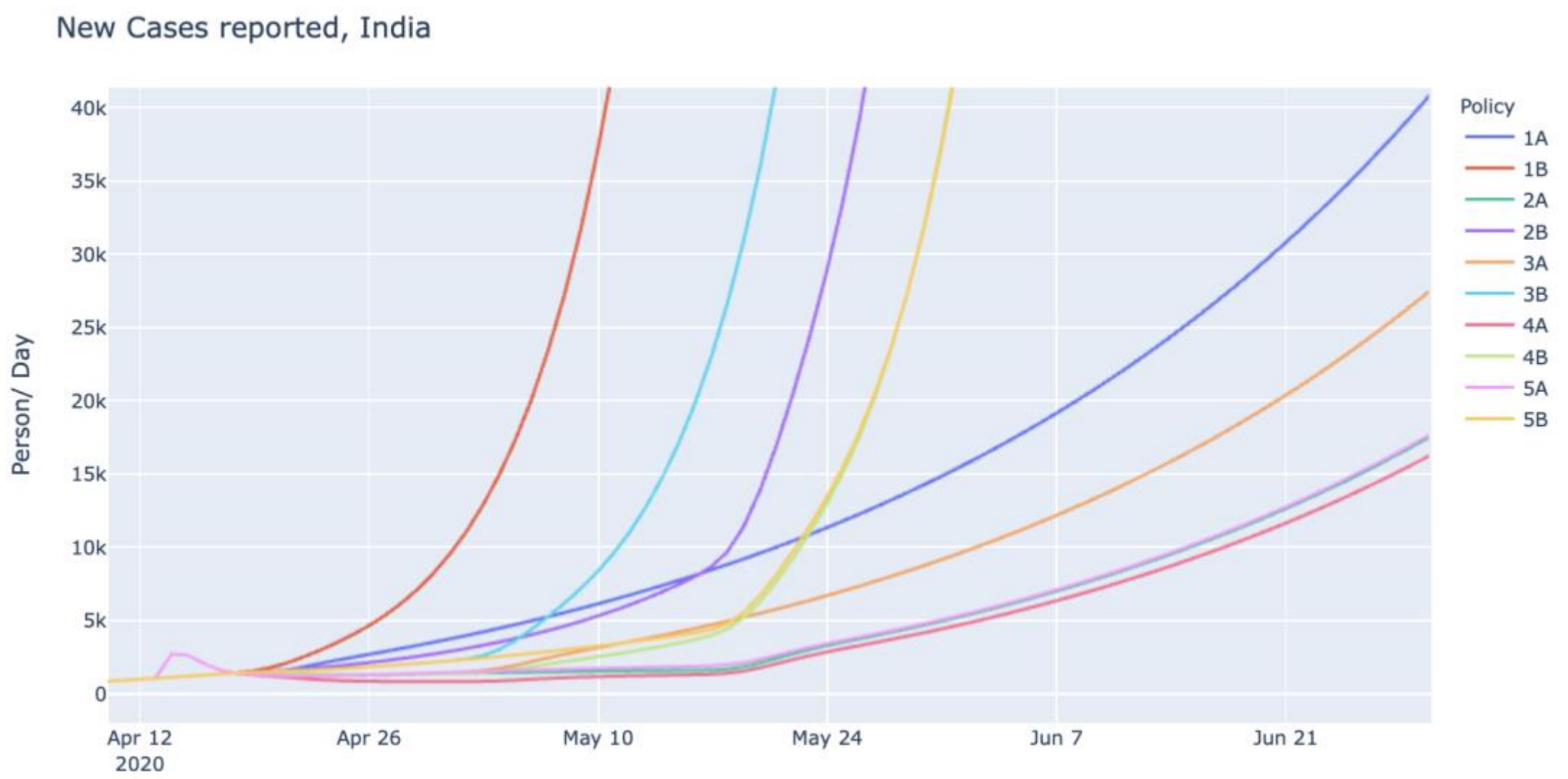

*Figure 6 Simulation response based on India data (until June 30th 2020), Zoomed in. The lower three lines are 4A, 2A and 5A*

The state-wise behaviour for polices 2A, 4A and 5A are show in Figures 7, 8 and 9 respectively. All three policies show a similar trends in behaviour. It is observed from Figures that, as a result of response either policy (2A, 4A or 5A), the pandemic is well controlled in all states. However, MH and DL (top two lines) do show a relatively larger increase that the other states, while in KA and KL (bottom two lines), the pandemic is completely under control. The projected behaviour for all the other states are somewhere in between these extremes.

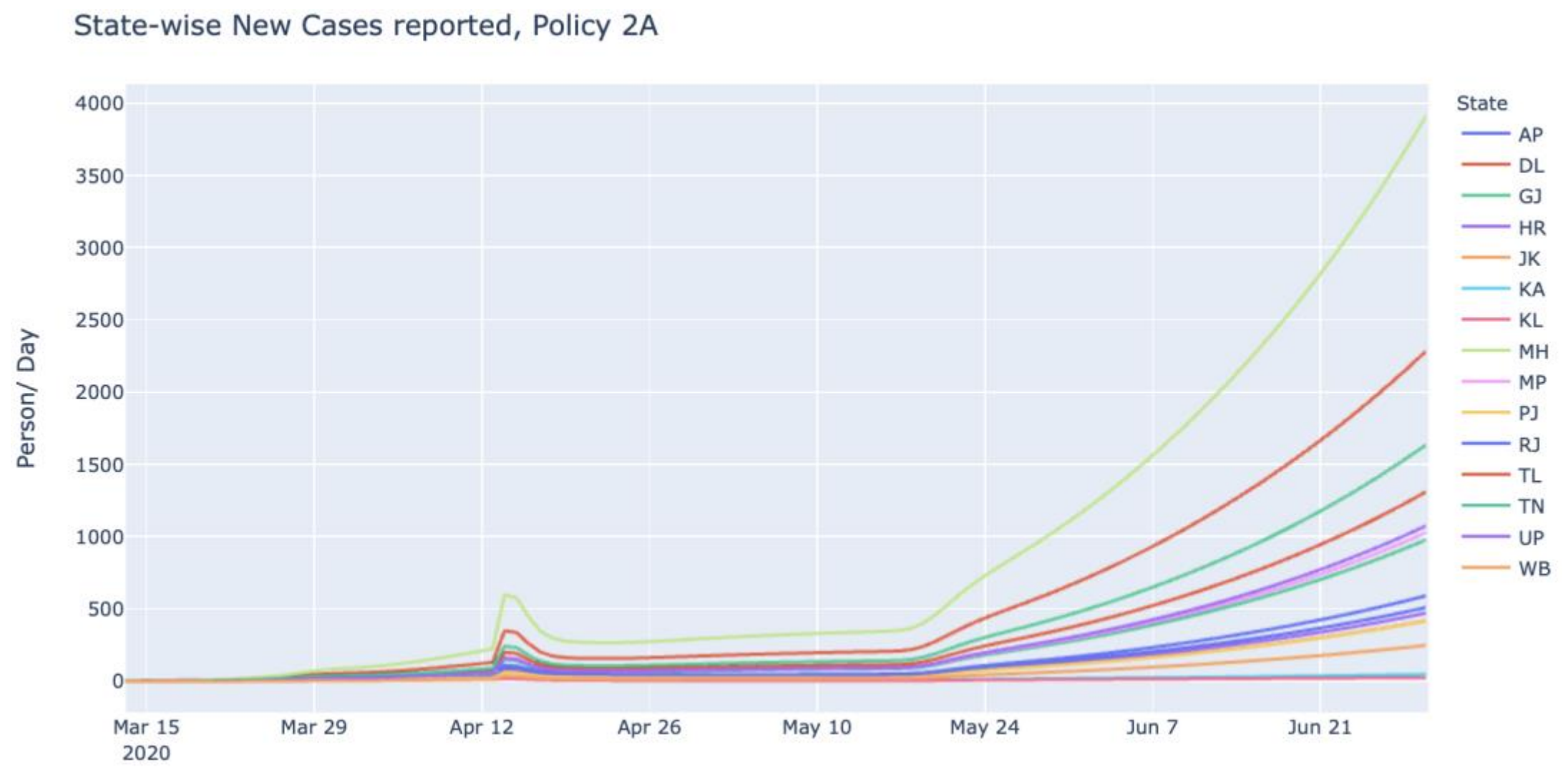

*Figure 7: Simulation output of state wise behavior for policy 2A (50% reduction in contacts till 15th May; and increased testing and tracing)*

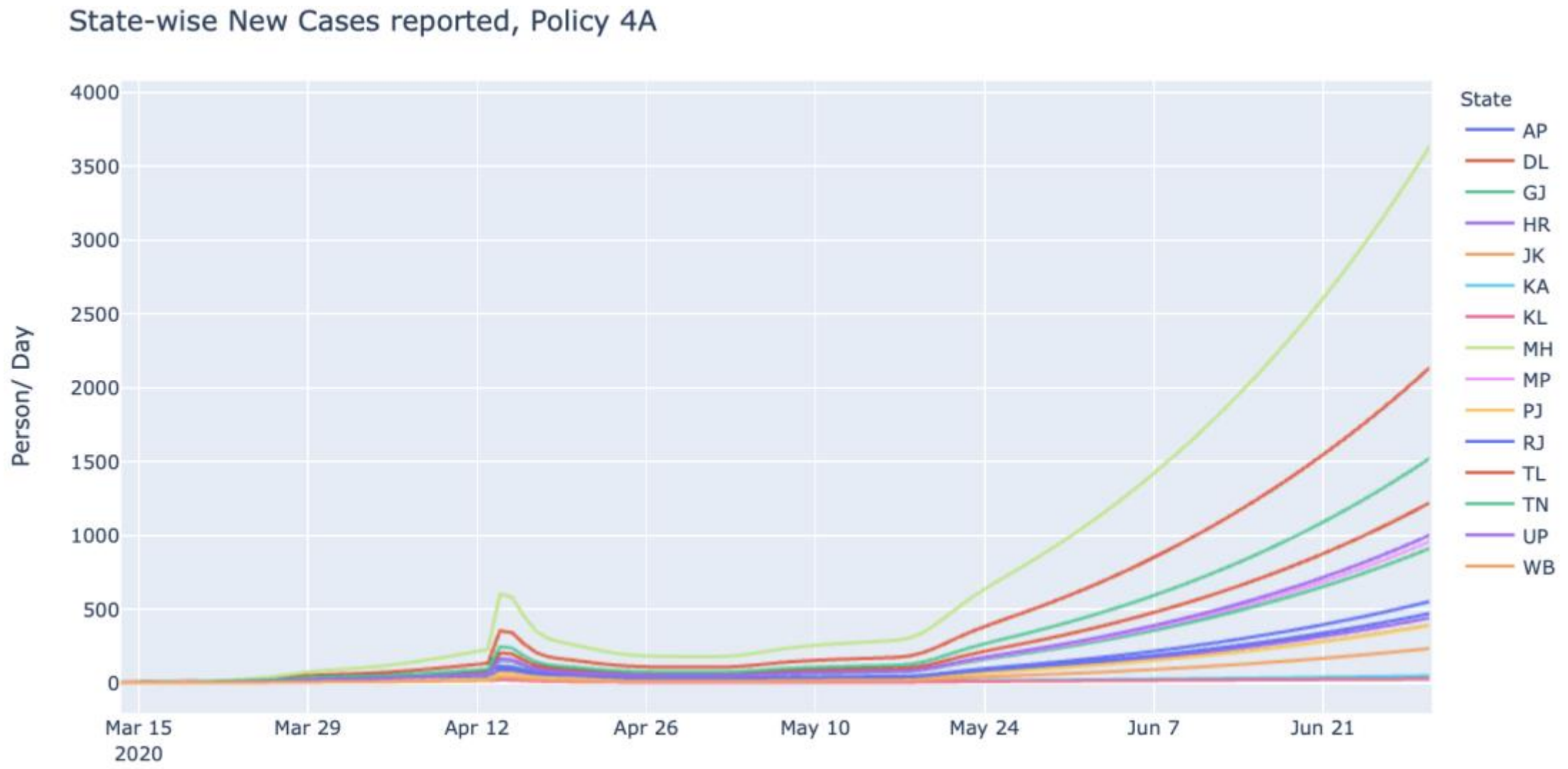

*Figure 8: Simulation output of state wise behavior for policy 4A (Extending the lockdown till 30th April and during 1st May to 15th May 50% reduction in contacts; and increased testing and tracing)*

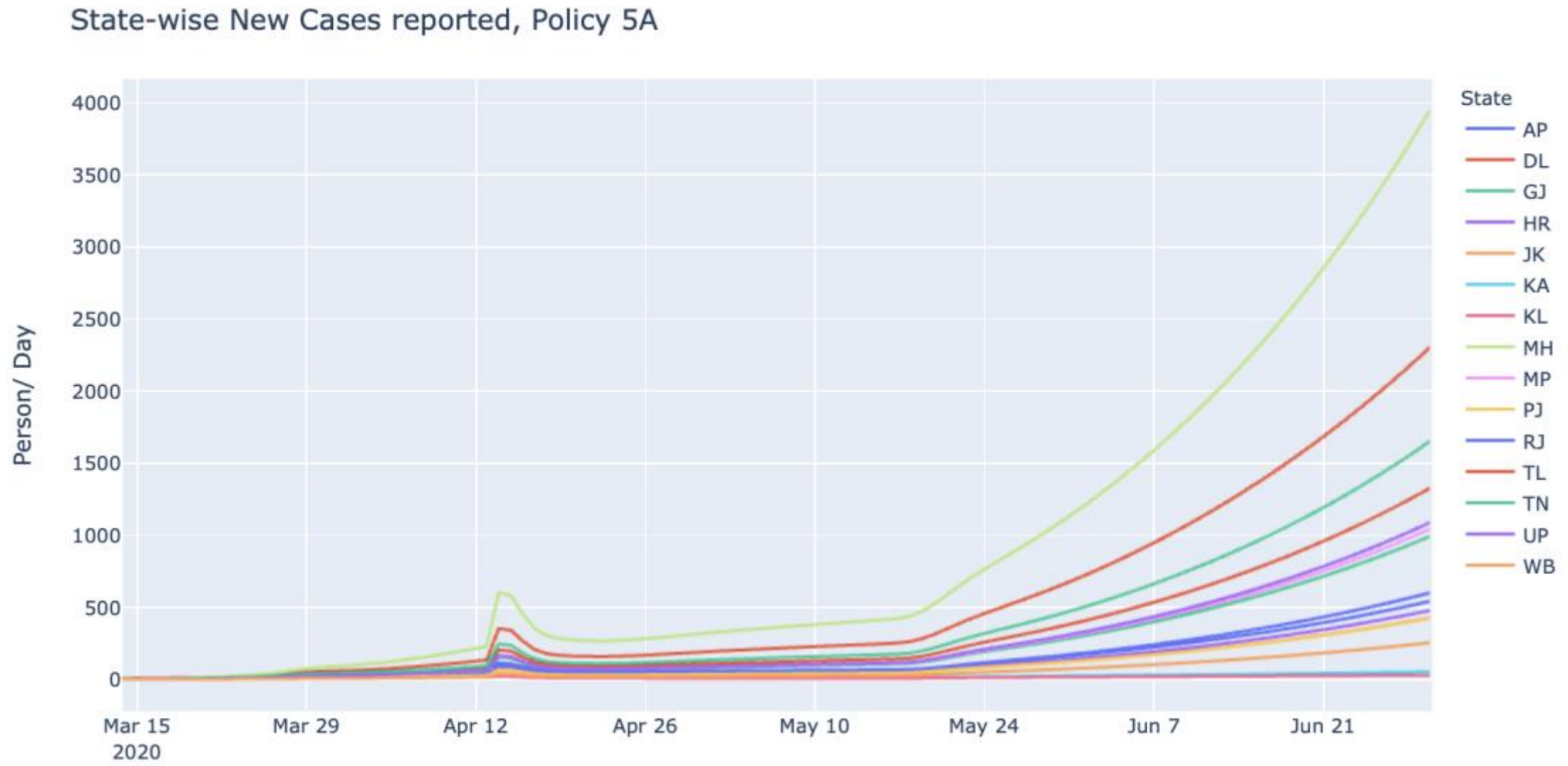

*Figure 9: Simulation output of state wise behavior for policy 5A (Extending the lockdown till May 15th; and increased testing and tracing)*

A few other policies have been evaluated (Policies 6A to 7D as mentioned in previous section). Figure 10 compares the behaviour of policy 6C, 6D, 7C and 7D with policy 4A (the best so far). It is readily apparent from the figure that policies 6C, 6D and (to some extent) 7C can help mitigate the pandemic spread. These show better projected behaviour than policy 4A. This shows that improved hygiene, social/ physical distancing, usage of masks and other protective gear can be quite effective. Unlike policy 6C (where Home, work and other contacts are at 50% and schools closed), in policy 7C only schools are closed, while home, work and others are back to 'normal'. However, the relative deterioration is only marginal between 6C and 7C (and still better than policy 4A). This again is due to improved hygiene, social/ physical distancing, usage of masks and other protective gear. The practical implementation of this policy remains to be seen. Figures 11 and 12 illustrate the state-wise response to policy 6C and 7C respectively. Policy 6C (see figure 11) is effective in readily controlling the pandemic spread. Although policy 7C (figure 12) shows an increasing rates, the growth is not exponential but asymptotic (which is a good thing, means the pandemic will not go out of control).

Do note that all the policies (such as increased testing/ contact tracing/ use of masks etc.) discussed in this section are assumed to be working in perpetuity.

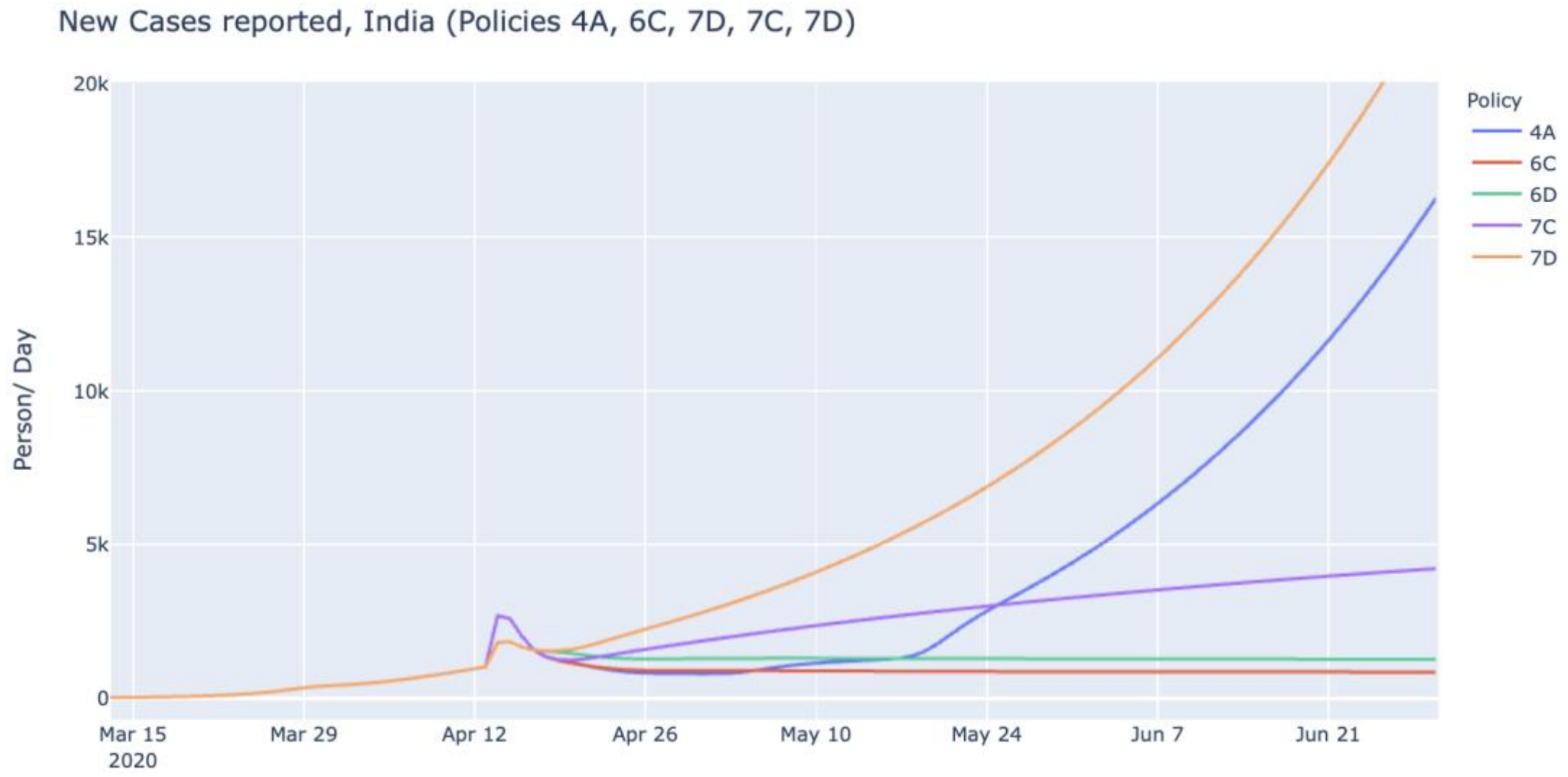

*Figure 10: Simulation output of all India behavior for other policies vs Policy 4A*

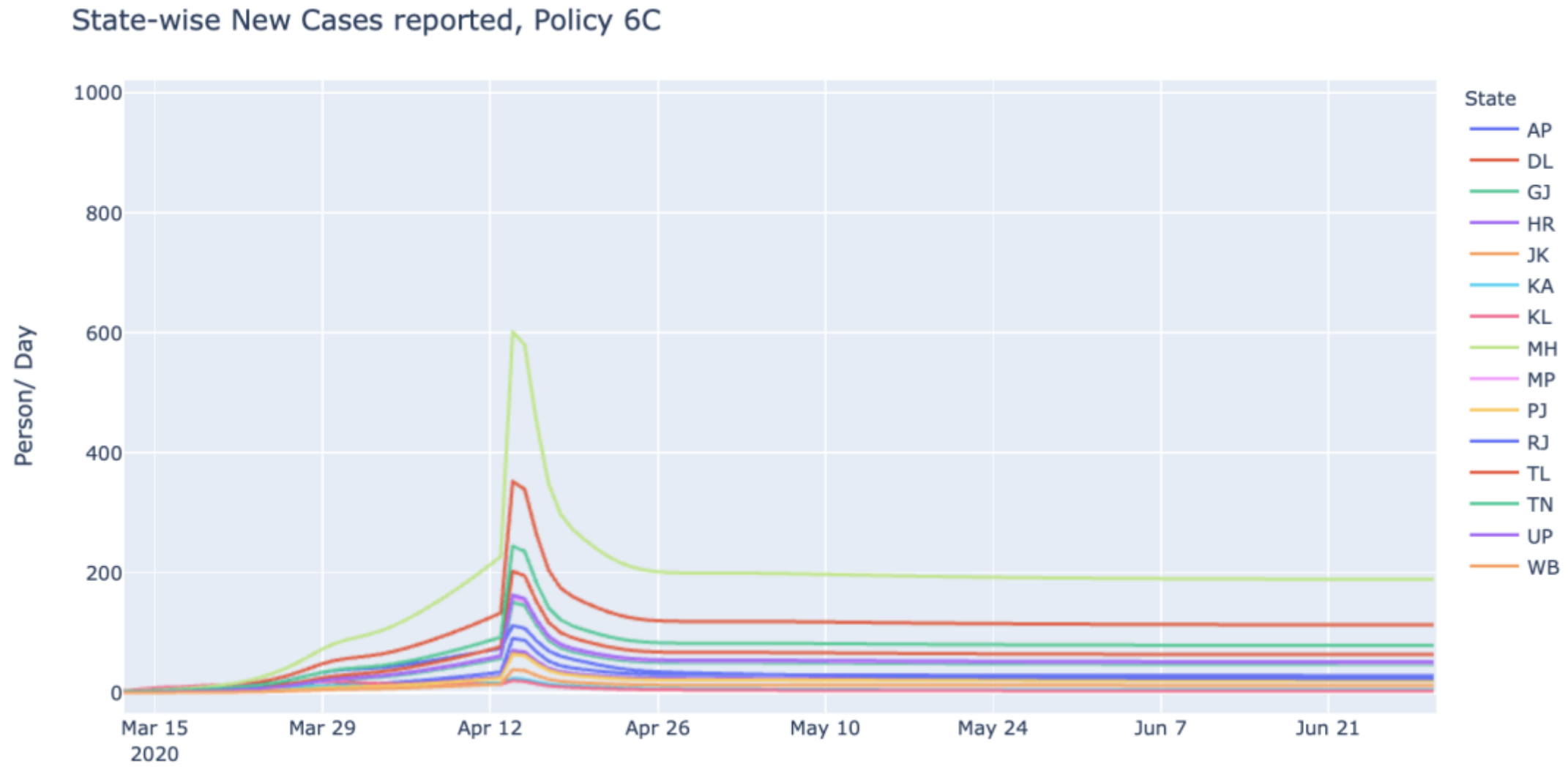

*Figure 11: Simulation output of state wise behavior for policy 6C (50% reduction in contacts, school closed, increased testing/ tracing; improved use of masks and better hygiene)*

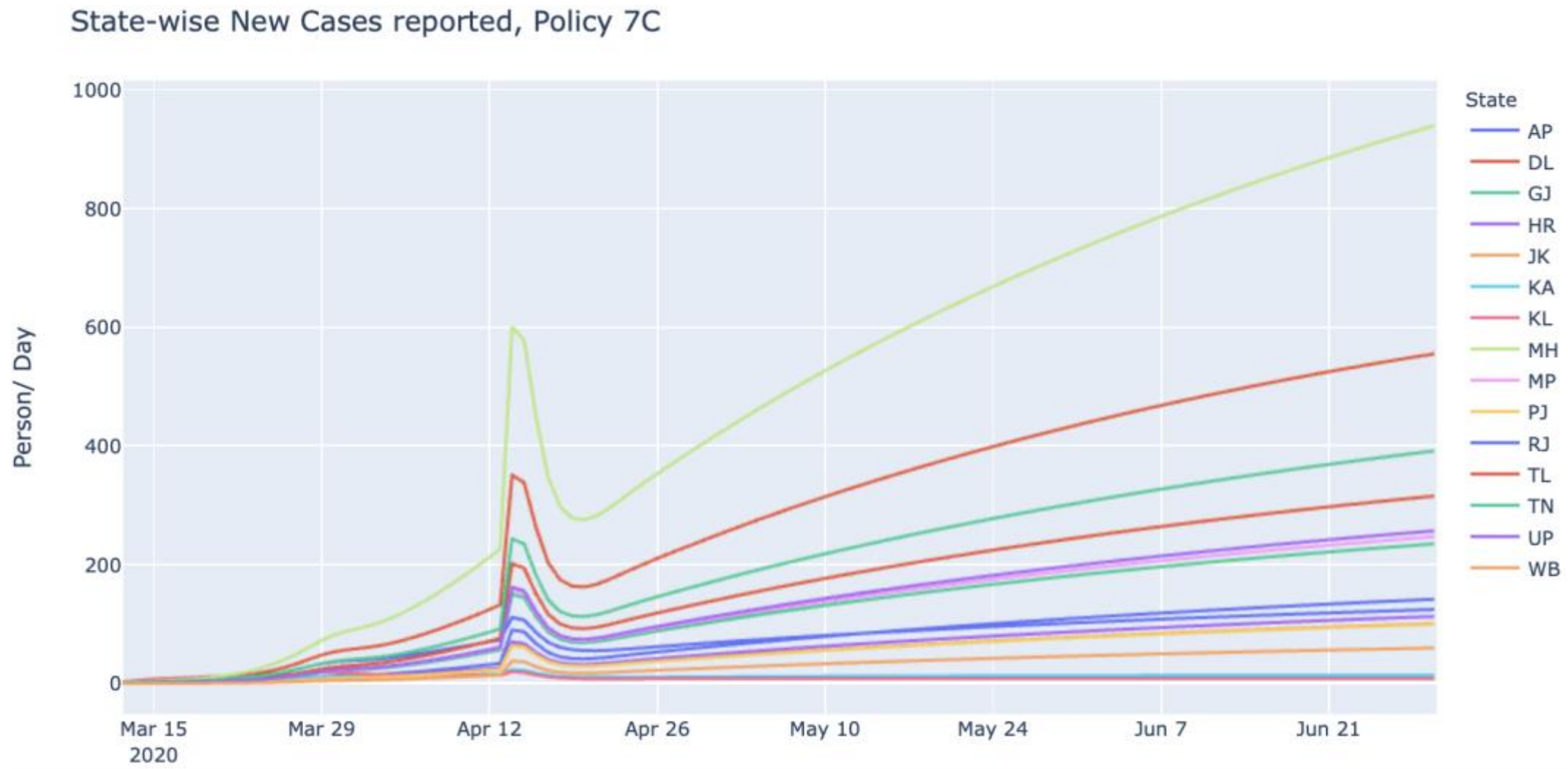

*Figure 12: Simulation output of state wise behavior for policy 7C (Only schools are closed, increased testing/ tracing; improved use of masks and better hygiene)*

## 9. Summary and the Way Forward

For the Covid-19 pandemic in India, contrary to (Singh and Adhikari, 2020), our model shows that, even after a 49-day lock-down, some non-trivial number of infections (even asymptomatic) will be left and the pandemic will resurface. Only tool that works against the pandemic is high rate of testing those who show Covid-19 like symptoms, isolating them if they are positive and contact tracing all contacts of positive patients and quarantining them. In combination with use of face masks and personal hygiene, contact tracing and isolation need not be super accurate. A wide range of combination of contact tracing, isolation, quarantining and personal hygiene measures keep the pandemic at bay and imperfections in one sphere are easily compensated by better implementation of other measures. This works if we keep just the schools closed for 90 days and allow all other activities with reasonable social distancing (only 20% improvement compared to base is sufficient). This works well if we open work but reduce non-essential contacts to half, while keeping schools closed for 90 days. Thus, we recommend that, during the lockdown, appropriate measures are put in place to do effective contract tracing, testing and isolations, increase the awareness & responsibility among the citizens to abide by social/ physical distancing norms, usage of masks and other PPE, better hygiene practices, reduced discretionary trip and improved sanitisation. This is will then help keep the pandemic in check while we can slowly reopen the economic activities.

A few immediate future works with regards to modelling, simulation and analysis are (not in any particular order):

- State-wise sensitivity analysis of various policy levers to help the level of adherence/ enforcement of each (say, testing, contact tracing, hygiene, usage of masks, sanitisation, contact rate reduction, physical distancing, etc.)
- Explore the impact of quarantining of symptomatics and asymptomatics though awareness efforts.
- Explore the impact of isolating asymptomatics though testing. Currently policy is to test only symptomatics.
- Use of patient case history to improving the model with respect to recovery delays, resource use, comorbidity, etc.
- The changes in $y(t)$ and $v(t)$ to be made endogenous, driven by feedback of the local reported cases and/or deaths.
- Understand and if possible, make explicit in the model the causes for policy resistance and unintended consequences.

The Vensim model can be shared on request.

## Appendix A: Model Notations and Equations

*Table 3: Notations used*

| Notation | Variable name | Description | *Units* |
|---|---|---|---|
| $S_i$ | Susceptibles | Population of age-group *i susceptible* to virus | *Person* |
| $E_i$ | Exposed | Population of age-group *i exposed* to virus, but not infective | *Person* |
| $A_i$ | Asymptomatics Infectives | Population of age-group *i* who are infected, show no symptoms, and are infective | *Person* |
| $I_i$ | Infectious Symptomatics | Population of age-group *i* who are infected and show symptoms (mild) | *Person* |
| $H_i$ | Hospitalised Patients | Population of age-group *i* who exhibit serious symptoms, tested positive, and are in hospital care | *Person* |
| $C_i$ | Critical Patients | Population of age-group *i* who exhibit extreme symptoms and may need ICU/ ventilator support | *Person* |
| $R_i$ | Recovered | Cumulative population of age-group *i* who have recovered | *Person* |
| $D_i$ | Dead | Cumulative population of age-group *i* who are dead due to the infection | *Person* |
| $Q_i^A$ | Quarantined Asymptomatics | Population of age-group *i* in quarantine who are infected, show no symptoms, and are infective | *Person* |
| $Q_i^I$ | Quarantined Symptomatics | Population of age-group *i* in quarantine who are infected and show symptoms | *Person* |
| $L_i^A$ | Isolated Asymptomatics | Population of age-group *i* in isolation who are infected, show no symptoms, are infective and tested positive. | *Person* |
| $L_i^I$ | Isolated Symptomatics | Population of age-group *i* in isolation who are infected, show symptoms and tested positive. | *Person* |
| $OH_i$ | Hospitalised Patients Overflow | Population of age-group *i* who exhibit serious symptoms, tested positive, and are in hospital care but overflow | *Person* |
| $OC_i$ | Critical Patients Overflow | Population of age-group *i* who exhibit extreme symptoms and may need ICU/ ventilator support but overflow | *Person* |
| $DH_i$ | Demand on Hospital | Population of age-group *i* who exhibit serious symptoms, tested positive, and requiring hospital care | *Person* |
| | | | |
| $IR_i(t)$ | Infection Rate | Rate of infection of Susceptible population | *Person/Day* |
| $ISR_i(t)$ | Infection Setting Rate | Rate at which exposed individuals become infectious | *Person/Day* |
| $XAR_i(t)$ | External arrivals of Asymptomtics | Rate at which external infected asymptomatics arrive in the community | *Person/Day* |
| $InR_i(t)$ | Incubation Rate | Rate at which asymptomatic individuals exhibit symptoms | *Person/Day* |
| $DPR_i(t)$ | Disease Progress Rate | Rate at which infectious symptomatics individuals becomes serious and move to hospital | *Person/Day* |
| $QAR_i(t)$ | Quarantining Asymptomatics Rate | Rate at which asymptomatic infectives are quarantined | *Person/Day* |
| $LAR_i(t)$ | Isolating Asymptomatics Rate | Rate at which asymptomatic infectives are isolated | *Person/Day* |
| $QSR_i(t)$ | Quarantining Symptomatics Rate | Rate at which Infectious Symptomatics are quarantined | *Person/Day* |
| $LSR_i(t)$ | Isolating Symptomatics Rate | Rate at which Infectious Symptomatics are tested positive and isolated | *Person/Day* |
| $QInR_i(t)$ | Quarantinee Incubation Rate | Rate at which Quarantined asymptomatic individuals exhibit symptoms | *Person/Day* |
| $LInR_i(t)$ | Isolatees Incubation Rate | Rate at which Isolated asymptomatic individuals exhibit symptoms | *Person/Day* |
| $QDPR_i(t)$ | Quarantinee disease progress Rate | Rate at which quarantined infectious symptomatics individuals becomes serious and move to hospital | *Person/Day* |
| $LDPR_i(t)$ | Isolatees disease progress Rate | Rate at which isolated infectious symptomatics individuals becomes serious and move to hospital | *Person/Day* |
| $WR_i(t)$ | Worsening Rate | Rate at which hospitalised patients become critical | *Person/Day* |
| $DR_i(t)$ | Dying Rate | Rate at which critical patients die | *Person/Day* |
| $HAR_i(t)$ | Hospital admissions Rate | Rate at which patients are hospitalised (proper) | *Person/Day* |
| $HAOR_i(t)$ | Hospital admissions Overflow Rate | Rate at which patients are hospitalised (overflow) | *Person/Day* |
| $CAR_i(t)$ | Critical admissions Rate | Rate at which critical patients are admitted for critical care (proper) | *Person/Day* |
| $CAOR_i(t)$ | Critical admissions Overflow Rate | Rate at which critical patients are admitted for critical care (overflow) | *Person/Day* |
| $OWR_i(t)$ | Overflow Worsening Rate | Rate at which overflow hospital patients become critical | *Person/Day* |
| $ODR_i(t)$ | Overflow Dying Rate | Rate at which overflow critical patients die | *Person/Day* |
| | | | |
| $ARR_i(t)$ | Asymptomatics Recovering Rate | Rate at which Asymptomatics recover | *Person/Day* |
| $SRR_i(t)$ | Symptomatics Recovering Rate | Rate at which Infectious Symptomatics recover | *Person/Day* |
| $QARR_i(t)$ | Quarantined Asymptomatics Recovering Rate | Rate at which Quarantined Asymptomatics recover | *Person/Day* |
| $LARR_i(t)$ | Isolated Asymptomatics Recovering Rate | Rate at which Isolated Asymptomatics recover | *Person/Day* |
| $QSRR_i(t)$ | Quarantined Symptomatics Recovering Rate | Rate at which Quarantined Symptomatics recover | *Person/Day* |
| $LSRR_i(t)$ | Isolated Symptomatics Recovering Rate | Rate at which Isolated Symptomatics recover | *Person/Day* |
| $HRR_i(t)$ | Hospitalised patients Recovering Rate | Rate at which Hospital patients recover | *Person/Day* |
| $CRR_i(t)$ | Critical patients Recovering Rate | Rate at which Critical care patients recover | *Person/Day* |
| $OHRR_i(t)$ | Overflow Hospital patients Recovering Rate | Rate at which overflow Hospital patients recover | *Person/Day* |
| $OCRR_i(t)$ | Overflow Critical patients Recovering Rate | Rate at which Overflow critical care patients recover | *Person/Day* |
| | | | |
| $d_{EA}$ | Average latency period | Average duration when exposed individuals do not show symptoms | *Day* |
| $d_{AI}$ | Average time to develop symptoms | Average duration for asymptomatic infected individuals to show symptoms. Note: Incubation period = $d_{EA} + d_{AI}$ | *Day* |
| $d_{\{AQ,AL,AR\}}$ | Average time for asymptomatics to {quarantine, isolate, recover} | Average duration for an asymptomatics infected individual to quarantine ($d_{AQ}$), isolate ($d_{AL}$) or recover ($d_{AR}$) | *Day* |
| $d_{\{QQ,QAR\}}$ | Average time for quarantined asymptomatics to {develop symptoms or recover} | Average duration for quarantined asymptomatics infected individual to develop symptoms ($d_{QQ}$), or recover ($d_{QAR}$) | *Day* |
| $d_{\{LL,LAR\}}$ | Average time for isolated asymptomatics to {develop symptoms or recover} | Average duration for an isolated asymptomatics infected individual to develop symptoms ($d_{LL}$), or recover ($d_{LAR}$) | *Day* |
| $d_{\{IH,IR\}}$ | Average time for infectious symptomatics to {be hospitalised or recover} | Average duration for infectious symptomatics individual to become serious and hospitalise ($d_{IH}$), or recover ($d_{IR}$) | *Day* |
| $d_{\{IQ,IL\}}$ | Average time for infectious symptomatics to be {quarantined or isolated} | Average duration for infectious symptomatics individual to be quarantined ($d_{IQ}$), or isolated after testing ($d_{IL}$). | *Day* |
| $d_{\{QH,QIR\}}$ | Average time for quarantined symptomatics to {hospitalise or recover} | Average duration for quarantined symptomatics individual to become serious and hospitalise ($d_{QH}$), or recover ($d_{QSR}$) | *Day* |
| $d_{\{LH,LIR\}}$ | Average time for isolated symptomatics to {hospitalise or recover} | Average duration for isolated symptomatics individual to become serious and hospitalise ($d_{LH}$), or recover ($d_{LSR}$) | *Day* |
| $d_{\{HC,HR\}}$ | Average time for hospitalised patient to {become critical or recover} | Average duration for hospitalised patients to become critical ($d_{HC}$), or recover ($d_{HR}$) | *Day* |
| $d_{\{CD,CR\}}$ | Average time for critical patients to {die or recover} | Average duration for critical patients to die($d_{CD}$), or recover ($d_{CR}$) | *Day* |
| | | | |
| $\lambda_i(t)$ | | Rate of infection of a susceptible individual | *1/Day* |
| $N_j$ | Total population of agegroup j | Total Population of age-group *j*. Note that $\sum_{j=1}^{M} N_j = 1$, where *M* is the number of age-groups | *Person* |
| $N$ | Total population | Total Population of the region | *Person* |
| $g_i$ | Population fraction | Fraction of population in agegroup *i* | *Dmnl* |
| $k$ | Reduction in infectiousness of asymptomatics | represents the infectivity reduction factor for asymptomatics. | *Dmnl* |
| $C_{ij}$ | Contact rate | represents the net contacts per day of age-group *i* has with age-group *j* in all locations of home, work, school and others. | *1/Day* |
| $C_{ij}^{Z}$ | Contact rate of quarantined/ isolated population group *Z* | represents the net contacts per day of age-group *i* has with age-group *j* in all locations of home, work, school and others, for quarantined asymptomatics (K=1), for isolated asymptomatics (K=2), quanrantined symptomatics (K=3), and isolated symptomatics (K=4) | *1/Day* |
| $u^{\{H,W,S,O\}}$ | Control factor for social distancing | Represents time dependent control measure imposed on contacts at home (H), work (W), school (S) and other (O) locations, respectively | *Dmnl* |
| $u^{Z,\{H,W,S,O\}}$ | Control factor for social distancing of quarantined/ isolated population group *Z* | Represents time dependent control measure imposed on contacts at home (H), work (W), school (S) and other (O) locations, respectively, for quarantined asymptomatics (Z=1), for isolated asymptomatics (Z=2), quanrantined symptomatics (Z=3), and isolated symptomatics (Z=4) | *Dmnl* |
| $p(t)$ | Infectivity | Net probability of a susceptible person getting infected with the virus in an interaction with an infectious symptomatic individual | *Dmnl* |
| $\hat{p}$ | Base infectivity | Base probability of a susceptible person getting infected with the virus in an interaction with an infectious symptomatic individual | *Dmnl* |
| $v(t)$ | Control factor for hygiene and PPE use | Represents time dependent control measure due to better hygiene and PPE use (e.g. masks), imposed on the infectivity | *Dmnl* |
| $f_i^S$ | Fraction developing symptoms | Fraction of asymptomatics developing sympotoms from age group *i* | *Dmnl* |
| $f_i^H$ | Fraction becoming serious | Fraction of Infectious Symptomatics becoming serious and get hospitalized, from age group *i* | *Dmnl* |
| $f_i^C$ | Fraction becoming critical | Fraction of hospitalized patients becoming critical and requiring ICU support, from age group *i* | *Dmnl* |
| $f_i^D$ | Fraction dying | Fraction of critical patients dying, from age group *i* | *Dmnl* |
| $f_i^{CO}$ | Fraction becoming critical (overflow) | Fraction of overflow hospitalized patients becoming critical and requiring ICU support, from age group *i* | *Dmnl* |
| $f_i^{DO}$ | Fraction dying (overflow) | Fraction of overflow critical patients dying, from age group *i* | *Dmnl* |
| $f_1(t)$ | Control fraction for awareness effort on asymptomatics | Represents the time dependent fraction of asymptomatics infected who are quarantined through awareness efforts. | *Dmnl* |
| $f_2(t)$ | Control fraction for testing of asymptomatics | Represents the time dependent fraction of asymptomatics infected who are tested positive and hence isolated. Note: $f_1(t) + f_2(t) < 0.9$ | *Dmnl* |
| $f_3(t)$ | Control fraction for awareness effort on Symptomatics | Represents the time dependent fraction of Symptomatics who are quarantined through awareness efforts. | *Dmnl* |
| $f_4(t)$ | Control fraction for testing of Symptomatics | Represents the time dependent fraction of Symptomatics who are tested positive and hence isolated. Note: $f_3(t) + f_4(t) < 0.9$ | *Dmnl* |
| $f_X(t)$ | Fraction for contact tracing of asymptomatics | Represents the time dependent fraction of Asymptomatics who are identified via contract tracing and hence quarantined. | *Dmnl* |
| $x(t)$ | Control fraction for asymptomatics traced per symptomatic | Represents the time dependent control fraction of Asymptomatics traced per fraction symptomatic tested. | *Dmnl* |
| $y(t)$ | Adjustment factor | Represents the time dependent net fraction increase in interactions | *Dmnl* |
| $DRC_i(t)$ | Daily reported cases | A tracking variable to capture the number of new infections reported daily | *Person/Day* |
| $CRC_i$ | Cumulative number of reported cases | A tracking variable to capture the cumulative number of new infections reported | *Person* |
| $K^H$ | Capacity of hospital (bed) | Number of hospital beds for treating Covid-19 patients (non-critical) | *Person* |
| $K^C$ | Capacity of Critical care unit | Number of critical care facility (Ventilators/ ICUs) for treating Covid-19 patients (non-critical) | *Person* |
| $\delta$ | Time step | Simulation timestep (currently set at 0.0625) | |

* Note: The units 'Dmnl' stands for Dimensionless.

## Equations underlying the SD model

Equations of the stocks

$$\frac{dS_i}{dt} = -IR_i(t) \quad (1)$$

$$\frac{dE_i}{dt} = IR_i(t) - ISR_i(t) \quad (2)$$

$$\frac{dA_i}{dt} = ISR_i(t) + XAR(t) - InR_i(t) - QAR_i(t) - LAR_i(t) - ARR_i(t) \quad (3)$$

$$\frac{dI_i}{dt} = InR_i(t) - DPR_i(t) - QSR_i(t) - LSR_i(t) - SRR_i(t) \quad (4)$$

$$\frac{dQ_i^A}{dt} = QAR_i(t) - QInR_i(t) - QARR_i(t) \quad (5)$$

$$\frac{dL_i^A}{dt} = LAR_i(t) - LInR_i(t) - LARR_i(t) \quad (6)$$

$$\frac{dQ_i^I}{dt} = QSR_i(t) + QInR_i(t) - QDPR_i(t) - QSRR_i(t) \quad 7$$

$$\frac{dL_i^I}{dt} = LSR_i(t) + LInR_i(t) - LDPR_i(t) - LSRR_i(t) \quad 8$$

$$\frac{dDH_i}{dt} = DPR_i(t) + QDPRQ_i(t) + LDPR_i(t) - HAR_i(t) \quad 9$$

$$\frac{dH_i}{dt} = HAR_i(t) - CAR_i(t) - CAOR_i(t) - HRR_i(t) \quad 10$$

$$\frac{dC_i}{dt} = CAR_i(t) - DR_i(t) - CRR_i(t) \quad 11$$

$$\frac{dD_i}{dt} = DR_i(t) + ODR_i(t) \quad 12$$

$$\frac{dOH_i}{dt} = HAOR_i(t) - OWR_i(t) - OHRR_i(t) \quad 13$$

$$\frac{dOC_i}{dt} = CAOR_i(t) + OWR_i(t) - ODR_i(t) - OCRR_i(t) \quad 14$$

$$\frac{dR_i}{dt} = ARR_i(t) + SRR_i(t) + QARR_i(t) + LARR_i(t) + QSRR_i(t) + QLRR_i(t) + HRR_i(t) + CRR_i(t) + OHRR_i(t) + OCRR_i(t) \quad 15$$

Equations underlying the Infection Rate (IR)

$$IR_i(t) = S_i(t) * p(t) * \lambda_i(t) \quad 16$$

$$\lambda_i(t) = \sum_{j=1}^{M} \left[ (1-k)\, C_{ij} \frac{A_j}{N_j} + C_{ij} \frac{I_j}{N_j} + C_{ij}^1 \frac{Q_j^A}{N_j} + C_{ij}^2 \frac{L_j^A}{N_j} + C_{ij}^3 \frac{Q_j^I}{N_j} + C_{ij}^4 \frac{L_j^I}{N_j} \right] \quad 17$$

$$C_{ij} = \left[ u^H(t) C_{ij}^H + u^W(t) C_{ij}^W + u^S(t) C_{ij}^S + u^O(t) C_{ij}^O \right] * y(t) \quad 18$$

$$C_{ij}^Z = \left[ u^{Z,H}(t) C_{ij}^H + u^{Z,W}(t) C_{ij}^W + u^{Z,S}(t) C_{ij}^S + u^{Z,O}(t) C_{ij}^O \right] * y(t), \quad \text{for } Z \epsilon \{1,2,3,4\} \quad 19$$

$$p(t) = v(t) * \hat{p} \quad 20$$

$$N_i = N * g_i \quad 21$$

Equations of the rates of flow

$$ISR_i(t) = DELAY3(IR_i(t), d_{EA}) \quad 22$$

$$QAR_i(t) = (f_1(t) + f_X(t) f_4(t)) * DELAY3(ISR_i(t) + EAR_i(t), d_{AQ}) \quad 23$$

$$LAR_i(t) = f_2(t) * DELAY3(ISR_i(t) + EAR_i(t), d_{AL}) \quad 24$$

$$InR_i(t) = f_i^S * (1 - f_1(t) - f_2(t) - f_X(t)) * DELAY3(ISR_i(t) + EAR_i(t), d_{AI}) \quad 25$$

$$ARR_i(t) = (1 - f_i^S) * (1 - f_1(t) - f_2(t) - f_X(t)) * DELAY3(ISR_i(t) + EAR_i(t),\ d_{AR}) \quad 26$$

$$QInR_i(t) = f_i^S * DELAY3(QAR_i(t), d_{QQ}) \quad (27)$$

$$QARR_i(t) = (1 - f_i^S) * DELAY3(QAR_i(t), d_{QAR}) \quad (28)$$

$$LInR_i(t) = f_i^S * DELAY3(LAR_i(t), d_{LL}) \quad (29)$$

$$LARR_i(t) = (1 - f_i^S) * DELAY3(LAR_i(t), d_{LAR}) \quad (30)$$

$$QSR_i(t) = f_3(t) * DELAY3(InR_i(t), d_{IQ}) \quad (31)$$

$$LSR_i(t) = f_4(t) * DELAY3(InR_i(t), d_{IL}) \quad (32)$$

$$DPR_i(t) = f_i^H * (1 - f_3(t) - f_4(t)) * DELAY3(InR_i(t), d_{IH}) \quad (33)$$

$$SRR_i(t) = (1 - f_i^H) * (1 - f_3(t) - f_4(t)) * DELAY3(InR_i(t), d_{IR}) \quad (34)$$

$$QDPR_i(t) = f_i^H * DELAY3(QSR_i(t) + QInR_i(t), d_{QH}) \quad (35)$$

$$QSRR_i(t) = (1 - f_i^H) * DELAY3(QSR_i(t) + QInR_i(t), d_{QIR}) \quad (36)$$

$$LDPR_i(t) = f_i^H * DELAY3(LSR_i(t) + LInR_i(t), d_{LH}) \quad (37)$$

$$LSRR_i(t) = (1 - f_i^H) * DELAY3(LSR_i(t) + LInR_i(t), d_{LIR}) \quad (38)$$

$$MHIR_i(t) = \frac{DH_i}{\sum_i DH_i} MAX\left(0, \left(K^H - \frac{\sum_i H_i}{\delta}\right)\right) \quad (39)$$

$$HAR_i(t) = MIN\left(\frac{DH_i}{\delta}, MHIR_i(t)\right) \quad (40)$$

$$HAOR_i(t) = MAX\left(0, \frac{DH_i}{\delta} - MHIR_i(t)\right) \quad (41)$$

$$WR_i(t) = f_i^C * DELAY3(HAR_i(t), d_{HC}) \quad (42)$$

$$HRR_i(t) = (1 - f_i^C) * DELAY3(HAR_i(t), d_{HR}) \quad (43)$$

$$MCIR_i(t) = \frac{WR_i}{\sum_i WR_i} MAX\left(0, \left(K^C - \frac{\sum_i C_i}{\delta}\right)\right) \quad (44)$$

$$CAR_i(t) = MIN\big(WR_i(t), MCIR_i(t)\big) \quad (45)$$

$$CAOR_i(t) = MAX\big(0, WR_i(t) - MCIR_i(t)\big) \quad (46)$$

$$OWR_i(t) = f_i^{CO} * DELAY3(HAOR_i(t), d_{HC}) \quad (47)$$

$$OHRR_i(t) = (1 - f_i^{CO}) * DELAY3(HAOR_i(t), d_{HR}) \quad (48)$$

$$DR_i(t) = f_i^D * DELAY3(\ WR_i(t), d_{CD}) \quad 49$$

$$CRR_i(t) = (1 - f_i^D) * DELAY3(WR_i(t), d_{CR}) \quad 50$$

$$ODR_i(t) = f_i^{DO} * DELAY3(\ CAOR_i(t) + OWR_i(t), d_{CD}) \quad 51$$

$$OCRR_i(t) = (1 - f_i^{DO}) * DELAY3(CAOR_i(t) + OWR_i(t), d_{CR}) \quad 52$$

$$DRC_i(t) = DPR_i(t) + LSR_i(t) + QDPR_i(t) + LAR_i(i\ ) \quad 53$$

$$\frac{dCRC_i}{dt} = DRC_i(t) \quad 54$$

$$f_X(t) = x(t) * f_4(t) \quad 55$$

Note: Delay3(_*input rate*_, *delay duration*_) represents a third-order material delay (DELAY3), which is a cascade of three first-order exponential delays, with the delay at each stage is one-third of the total *delay duration*.

## APPENDIX B: Contact Matrices and related notes.

We use the Social Contact Matrix from (Prem et. al. 2017). However, what we really need is the number of interactions per day and not number of contacts and hence we calibrate the model and apply a correction multiplier of 1.5 to all the base contact numbers for all four spheres viz Home, Work School, and Other given there. Note that this factor of 1.5 is also an artificial result of the fact that we take base infectivity to be .015 from (Singh and Adhikari 2020). Any change in the base infectivity number will have a corresponding impact on the base contact correction multipliers.

## APPENDIX C: Results of State-wise model calibration

State-wise calibration of the model. Only the plot of daily reported cases, and cumulative reported cases given here. The blue lines represent the simulation results, and red line shows the actual reported cases.

During lockdown, $u^H(t)$ = 1, $u^W(t)$=0.3, $u^S(t)$=0, $u^O(t)$ =0.3, and 1 otherwise for all states.
Some observations:

- Parameter settings for AP, GJ, JK, MP, MH, PJ, TL, and WB are same as All India.
- Parameter settings for DL, HR, TN are similar to each other; differs from All India only in the starting value of y(t).
- In KA, KL, and RJ the testing fractions and contact tracing, $f_4(t) = x(t) = 0.5$. This is reasonable since Kerala model and Bilwara models haves been much praised for their effective containment.

<table>
<tr><td>Andhra Pradesh</td><td></td><td></td></tr>
<tr><td>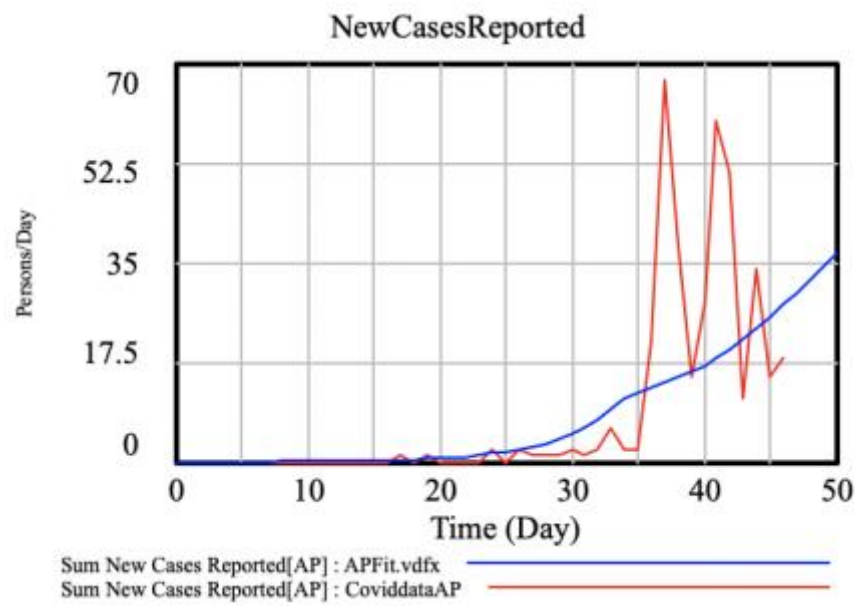
</td><td>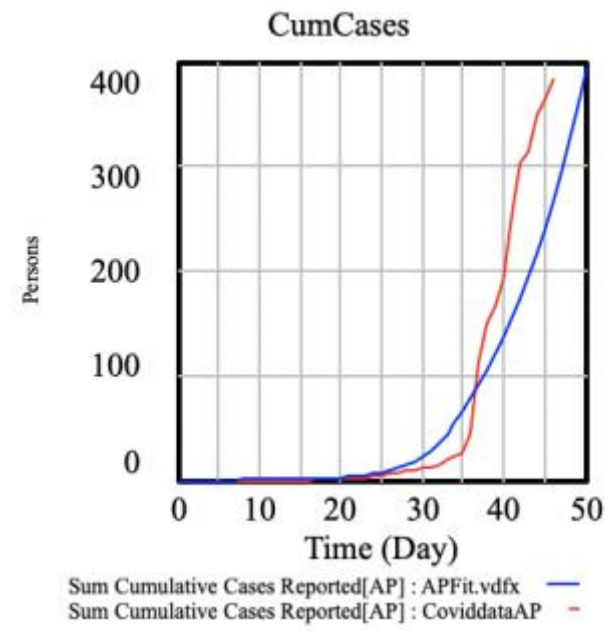
</td><td>$f_4(t)$ = 0.25<br>$x(t)$ = 0.1<br>$y(t) = \begin{cases} 2 & t \leq 15 \\ 2 - (\frac{0.8}{35}(t-15)) & 15 < t < 50 \end{cases}$</td></tr>
<tr><td>Delhi</td><td></td><td></td></tr>
<tr><td>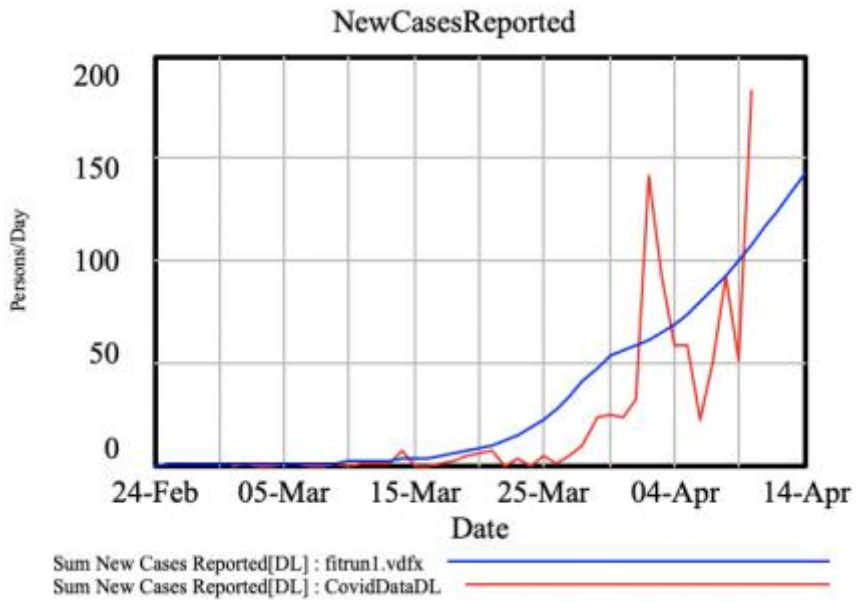
</td><td>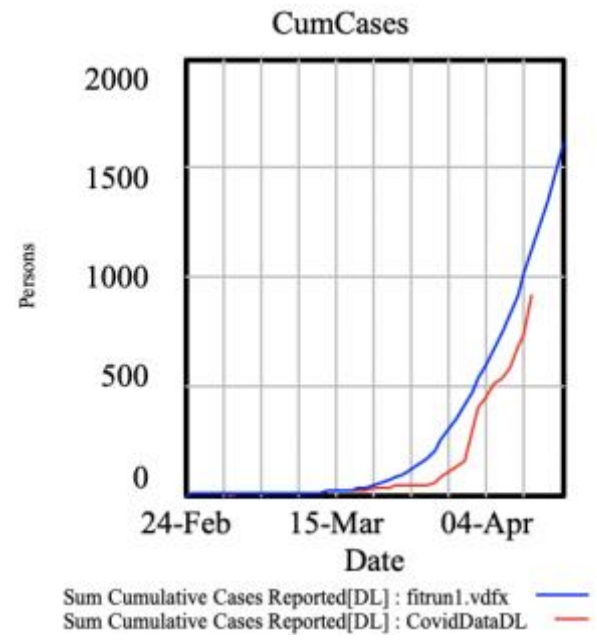
</td><td>$f_4(t)$ = 0.25<br>$x(t)$ = 0.1<br>$y(t) = \begin{cases} 1.6 & t \leq 15 \\ 1.6 - (\frac{0.4}{35}(t-15)) & 15 < t < 50 \end{cases}$</td></tr>
<tr><td>Gujarat</td><td></td><td></td></tr>
<tr><td>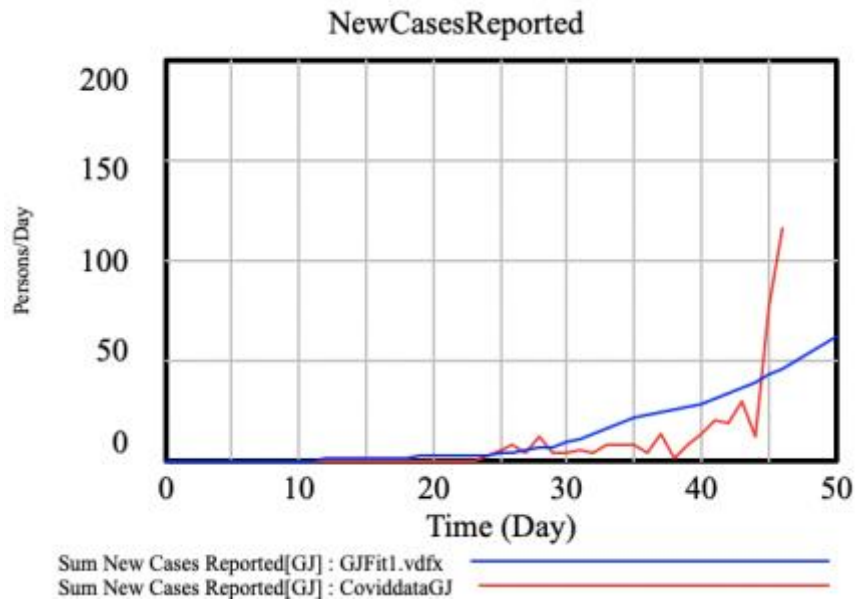
</td><td>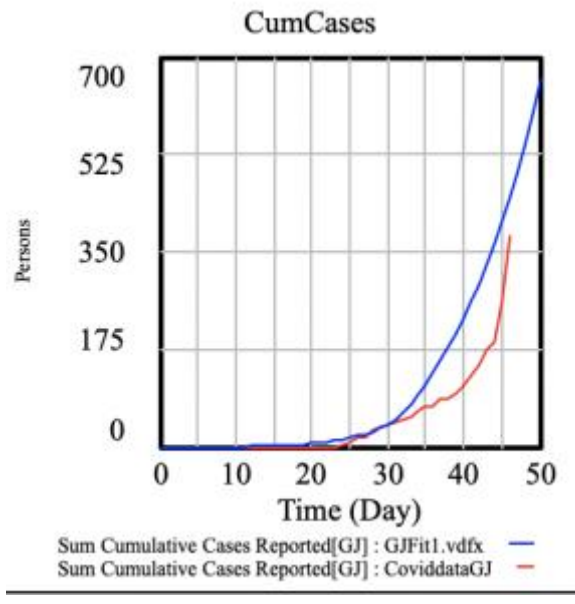
</td><td>$f_4(t)$ = 0.25<br>$x(t)$ = 0.1<br>$y(t) = \begin{cases} 2 & t \leq 15 \\ 2 - (\frac{0.8}{35}(t-15)) & 15 < t < 50 \end{cases}$</td></tr>
</table>

| | | |
|---|---|---|
| Haryana | | |
| 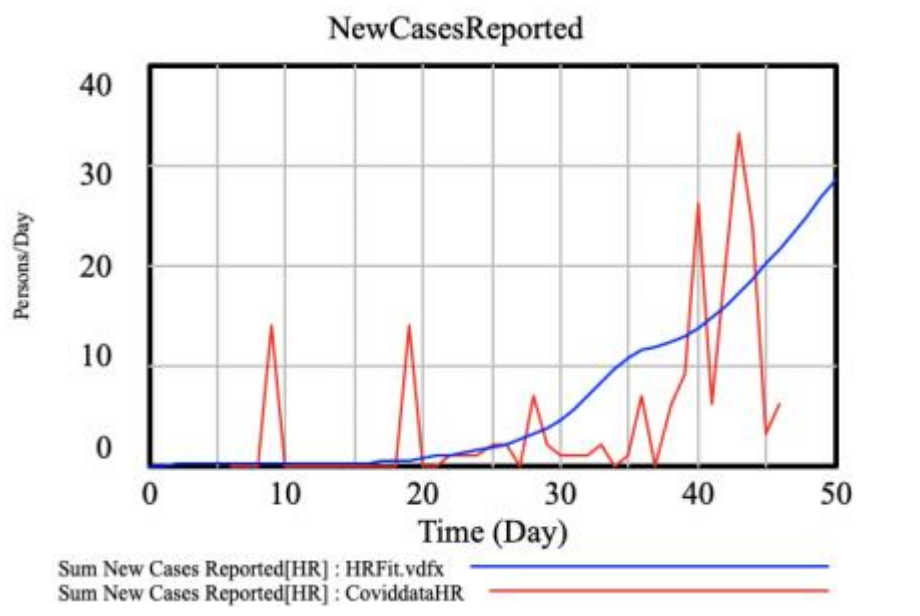 | 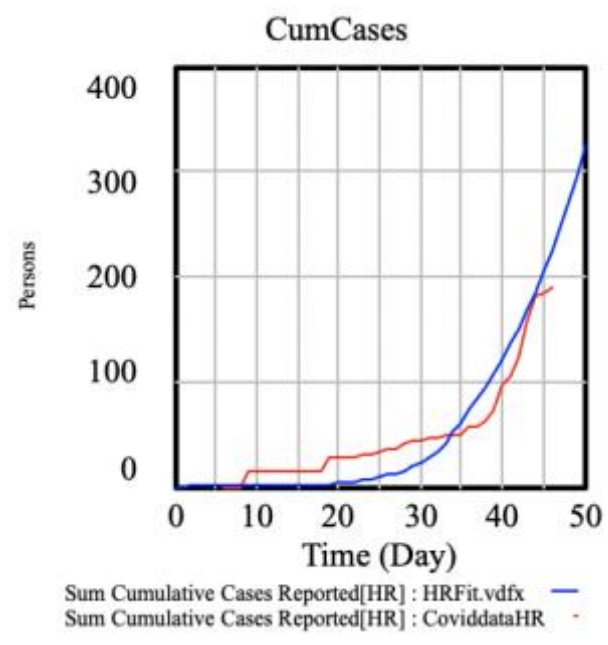 | $f_4(t) = 0.25$<br>$x(t) = 0.1$<br>$y(t) = \begin{cases} 1.6 & t \leq 15 \\ 1.6 - (\frac{0.4}{35}(t-15)) & 15 < t < 50 \end{cases}$ |
| Jammu and Kashmir | | |
| 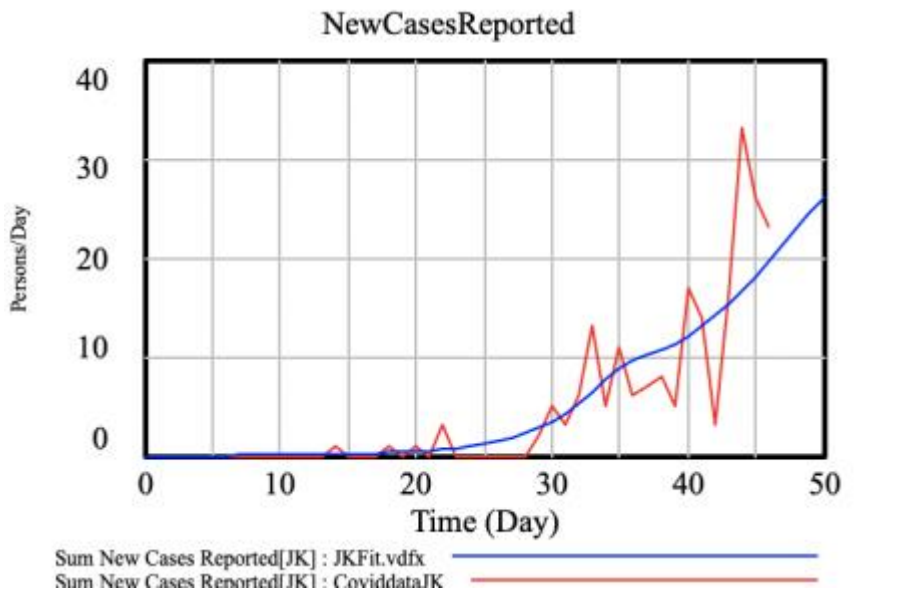 | 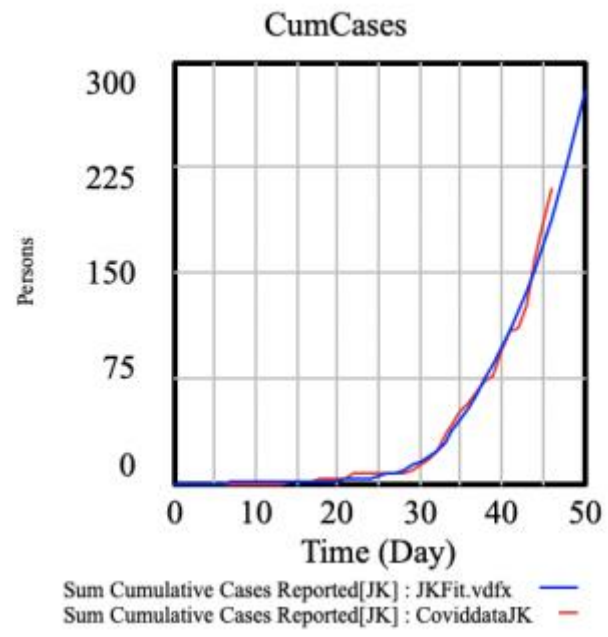 | $f_4(t) = 0.25$<br>$x(t) = 0.1$<br>$y(t) = \begin{cases} 2 & t \leq 15 \\ 2 - (\frac{0.8}{35}(t-15)) & 15 < t < 50 \end{cases}$ |
| Karnataka | | |
| 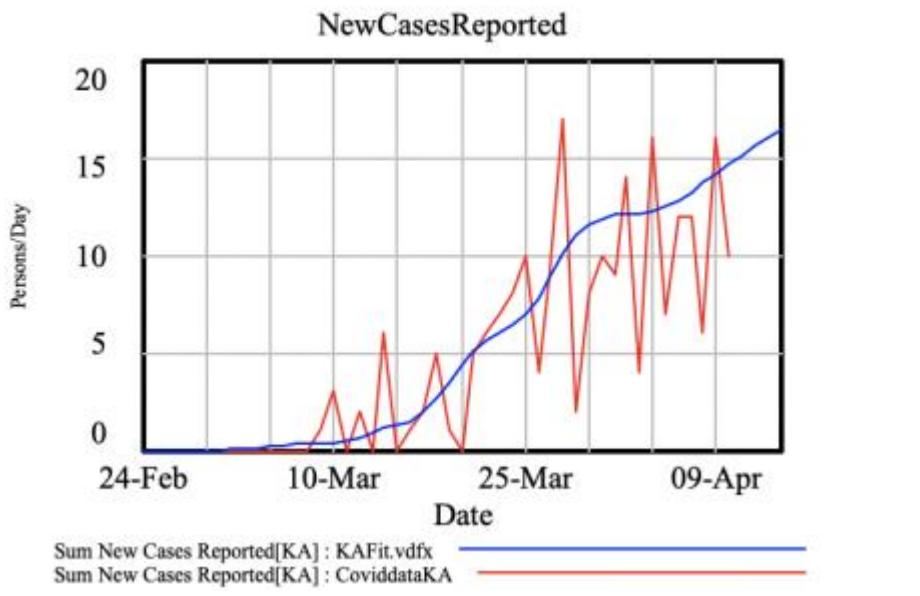 | 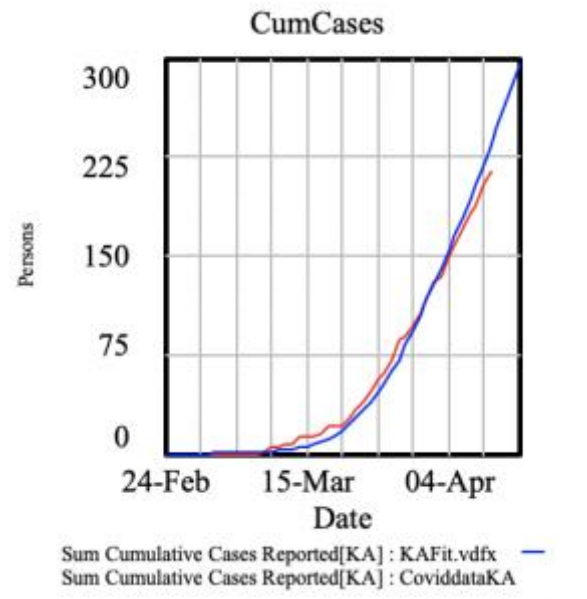 | $f_4(t) = 0.5$<br>$x(t) = 0.5$<br>$y(t) = \begin{cases} 1.5 & t \leq 15 \\ 1.5 - (\frac{0.5}{35}(t-15)) & 15 < t < 50 \end{cases}$ |
| Kerala | | |
| 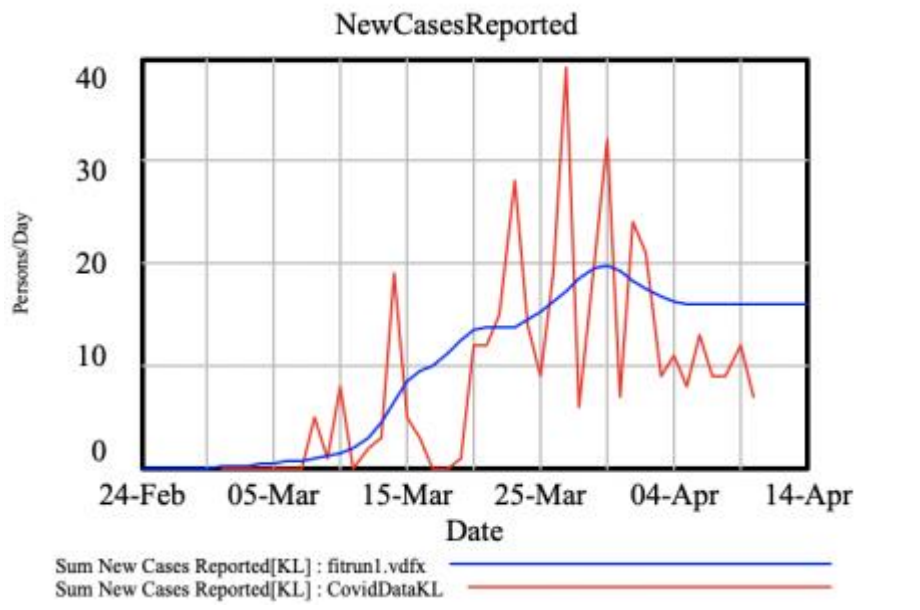 | 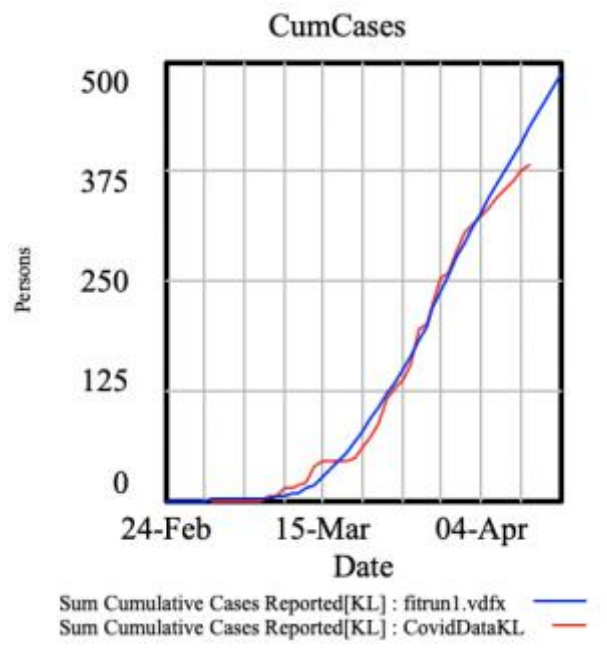 | $f_4(t) = 0.6$<br>$x(t) = 0.5$<br>$y(t) = \begin{cases} 1.5 & t \leq 5 \\ 1.5 - (\frac{0.5}{35}(t-15)) & 5 < t < 40 \end{cases}$<br>$v(t) = 0.8, t \geq 0$ |

<table>
<tr><td>Madhya Pradesh</td><td></td><td></td></tr>
<tr><td>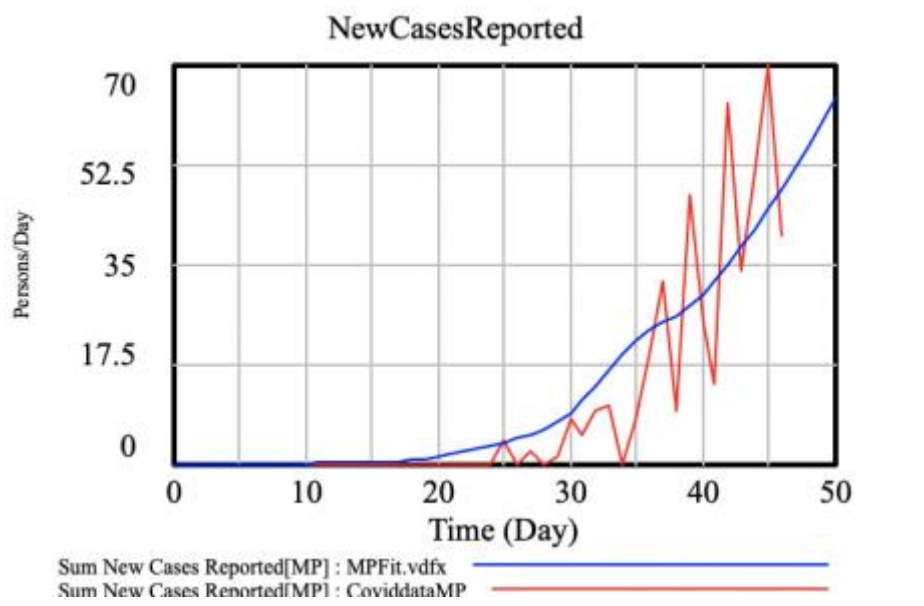
</td><td>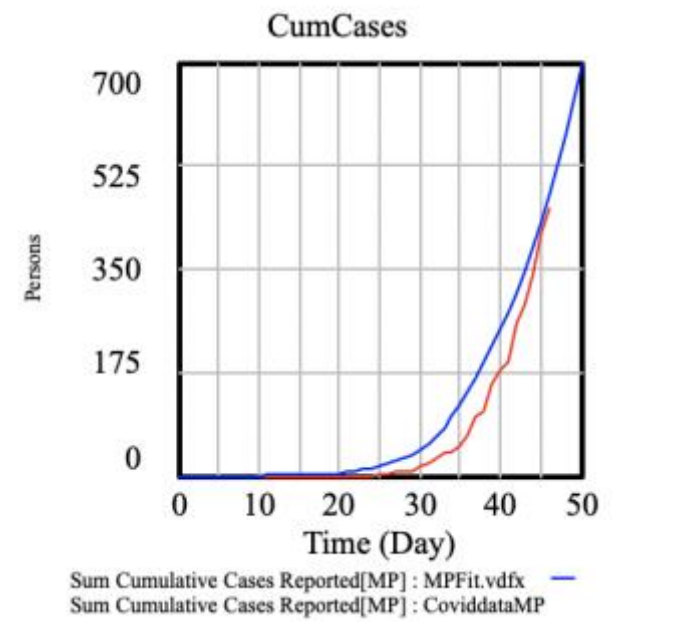
</td><td>$f_4(t) = 0.25$<br><br>$x(t) = 0.1$<br><br>$$y(t) = \begin{cases} 2 & t \le 15 \\ 2 - (\frac{0.8}{35}(t-15)) & 15 < t < 50 \end{cases}$$</td></tr>
<tr><td>Maharashtra</td><td></td><td></td></tr>
<tr><td>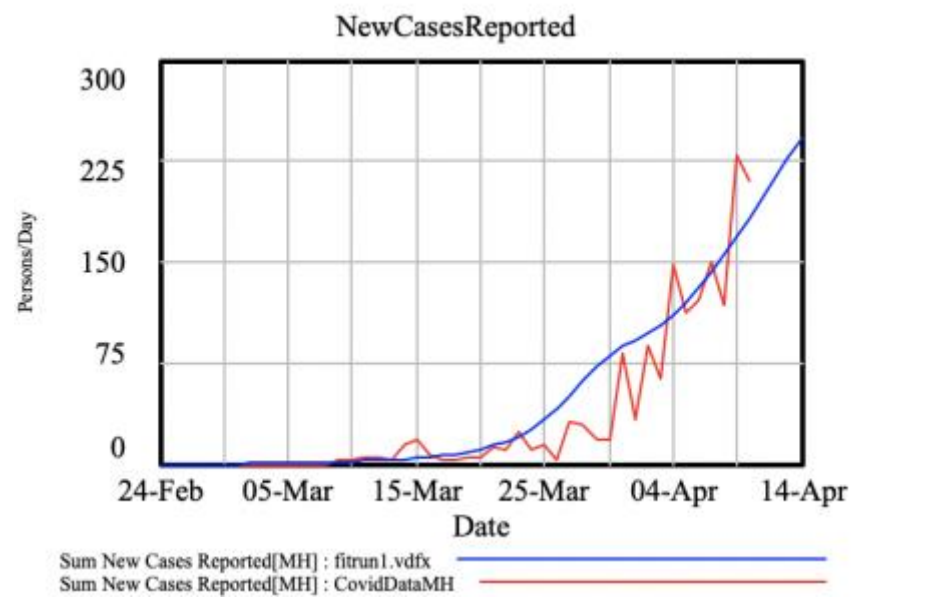
</td><td>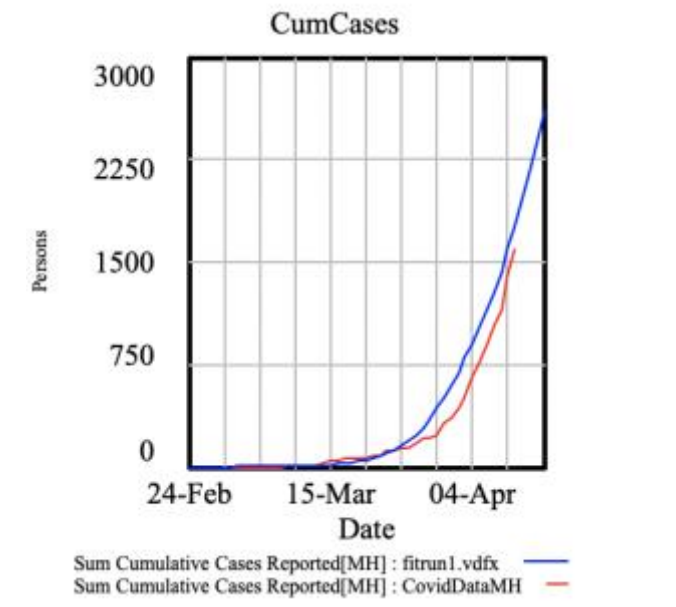
</td><td>$f_4(t) = 0.25$<br><br>$x(t) = 0.1$<br><br>$$y(t) = \begin{cases} 2 & t \le 15 \\ 2 - (\frac{0.8}{35}(t-15)) & 15 < t < 50 \end{cases}$$</td></tr>
<tr><td>Punjab</td><td></td><td></td></tr>
<tr><td>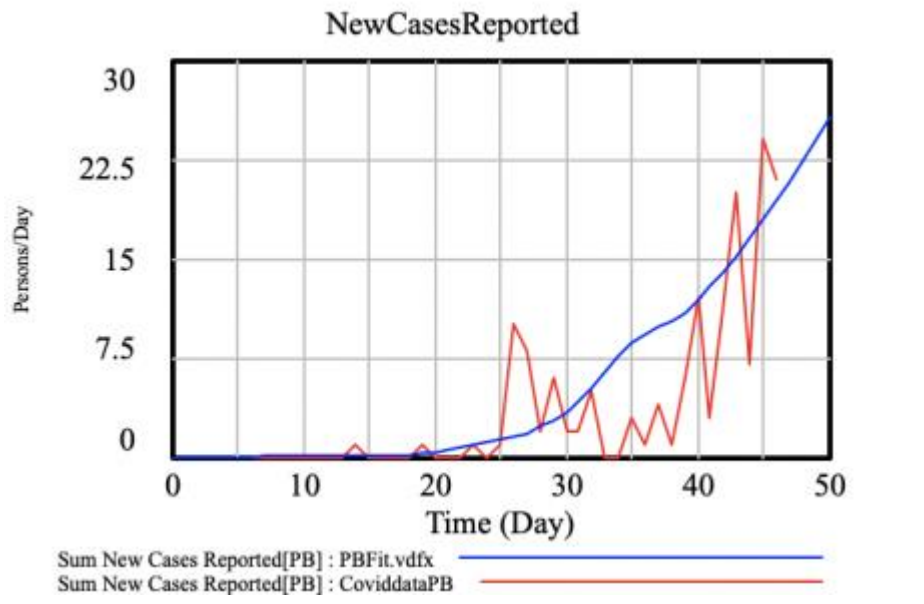
</td><td>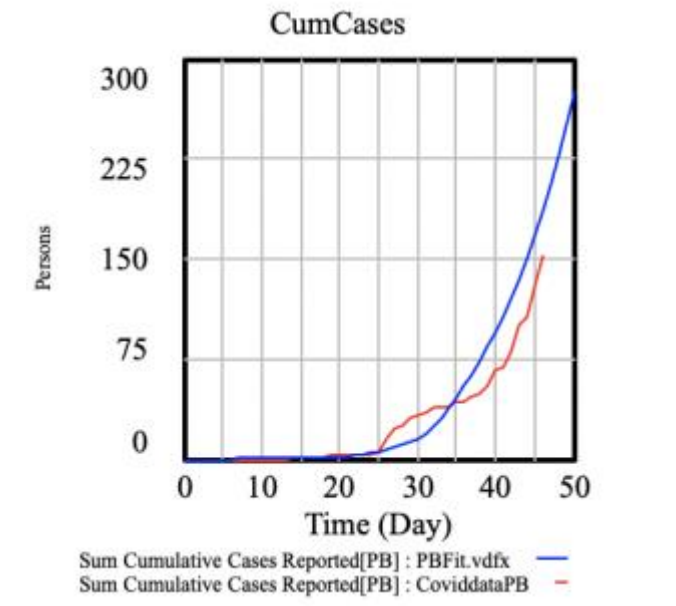
</td><td>$f_4(t) = 0.25$<br><br>$x(t) = 0.1$<br><br>$$y(t) = \begin{cases} 2 & t \le 15 \\ 2 - (\frac{0.8}{35}(t-15)) & 15 < t < 50 \end{cases}$$</td></tr>
<tr><td>Rajasthan</td><td></td><td></td></tr>
<tr><td>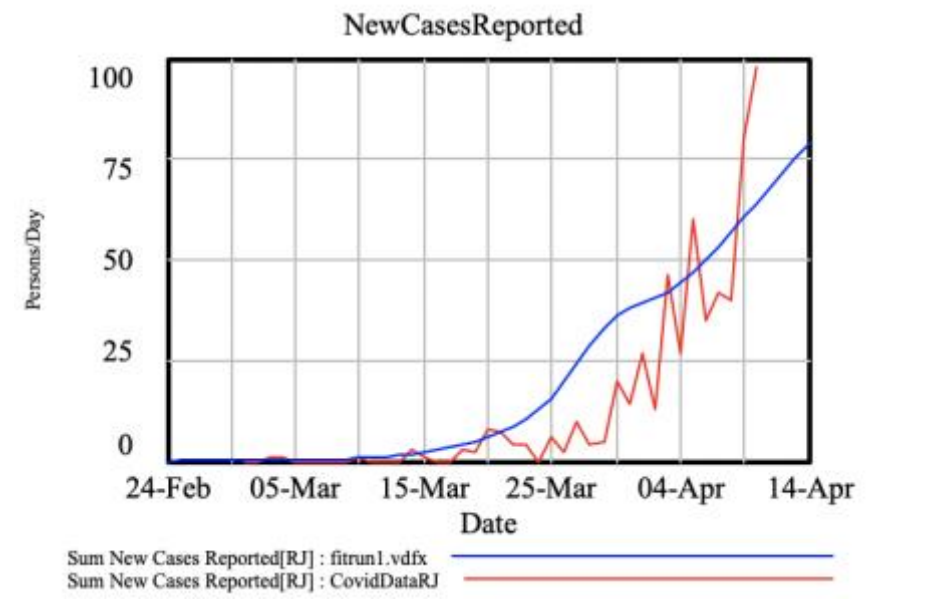
</td><td>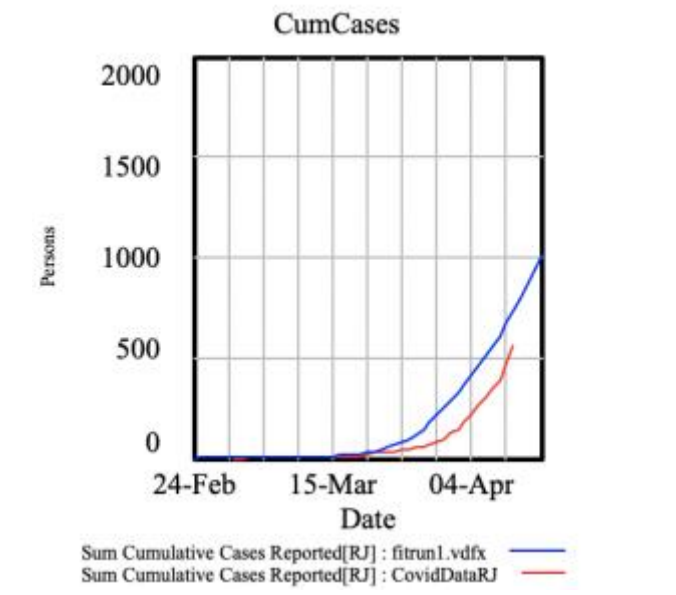
</td><td>$f_4(t) = 0.5$<br><br>$x(t) = 0.5$<br><br>$$y(t) = \begin{cases} 2 & t \le 15 \\ 2 - (\frac{0.8}{35}(t-15)) & 15 < t < 50 \end{cases}$$</td></tr>
</table>

<table>
<tr><td>Tamil Nadu</td><td></td><td></td></tr>
<tr><td>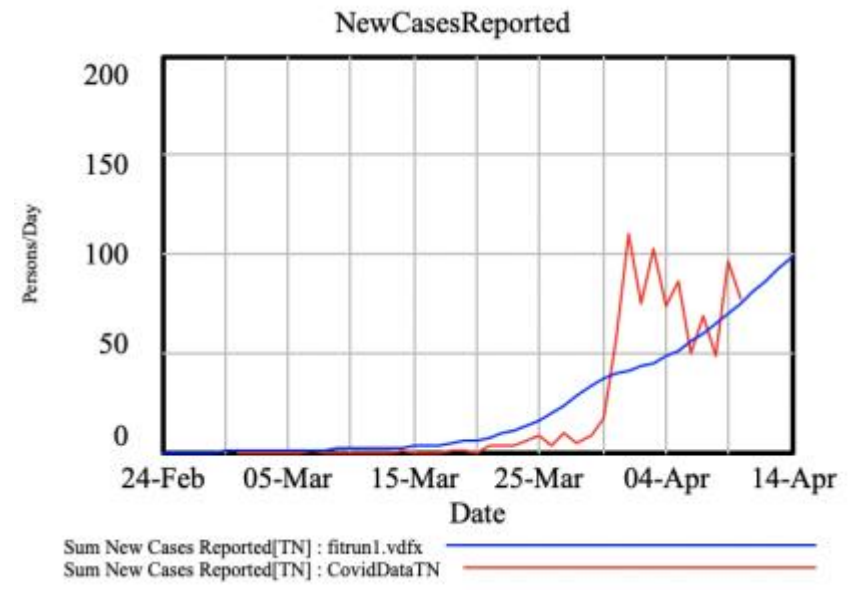
</td><td>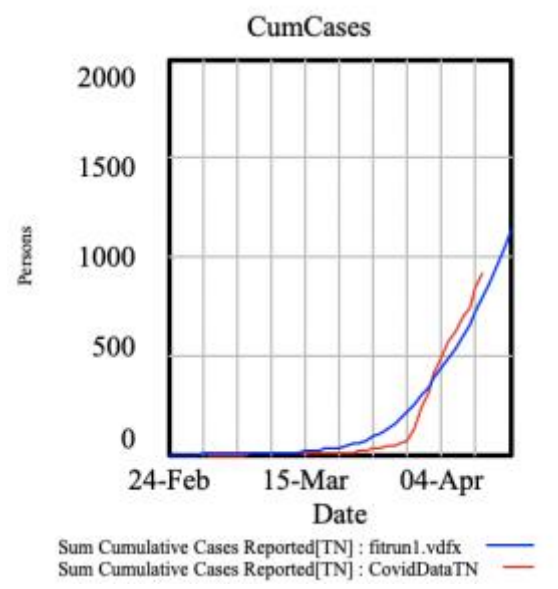
</td><td>$f_4(t) = 0.25$<br><br>$x(t) = 0.1$<br><br>$y(t) = \begin{cases} 1.6 & t \leq 15 \\ 1.6 - (\frac{0.4}{35}(t-15)) & 15 < t < 50 \end{cases}$</td></tr>
<tr><td>Telangana</td><td></td><td></td></tr>
<tr><td>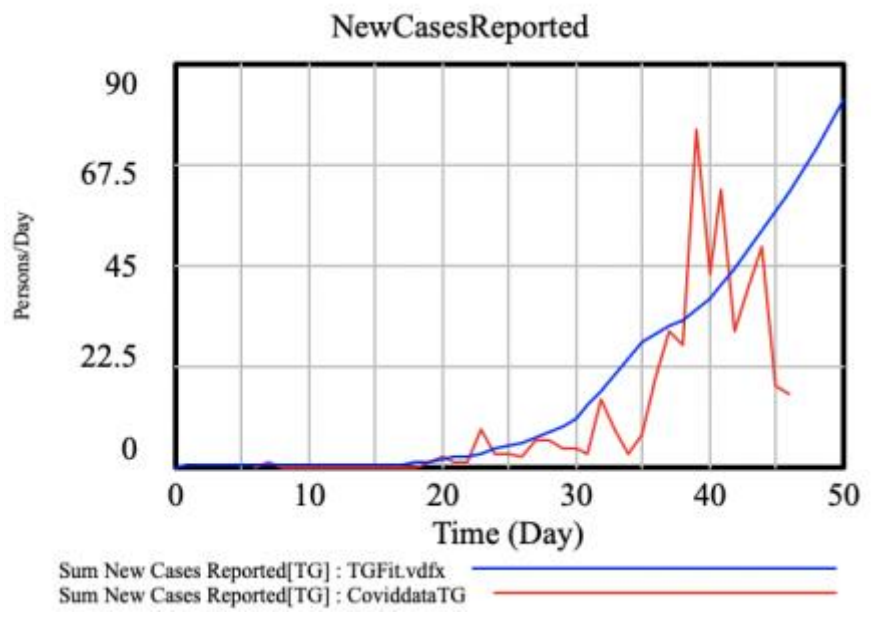
</td><td>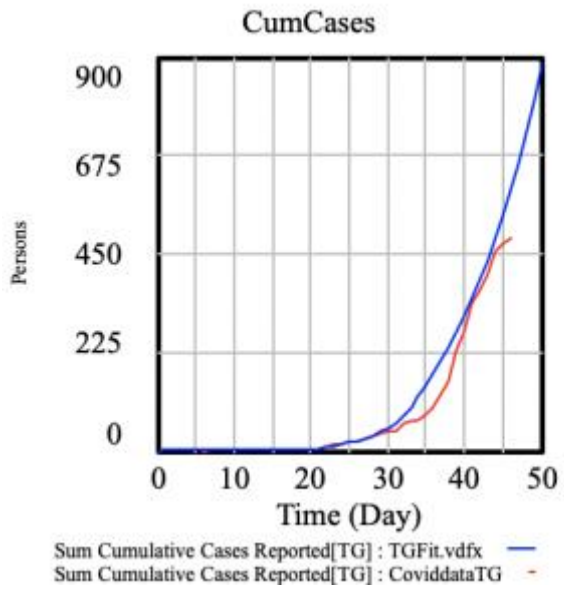
</td><td>$f_4(t) = 0.25$<br><br>$x(t) = 0.1$<br><br>$y(t) = \begin{cases} 2 & t \leq 15 \\ 2 - (\frac{0.8}{35}(t-15)) & 15 < t < 50 \end{cases}$</td></tr>
<tr><td>Uttar Pradesh</td><td></td><td></td></tr>
<tr><td>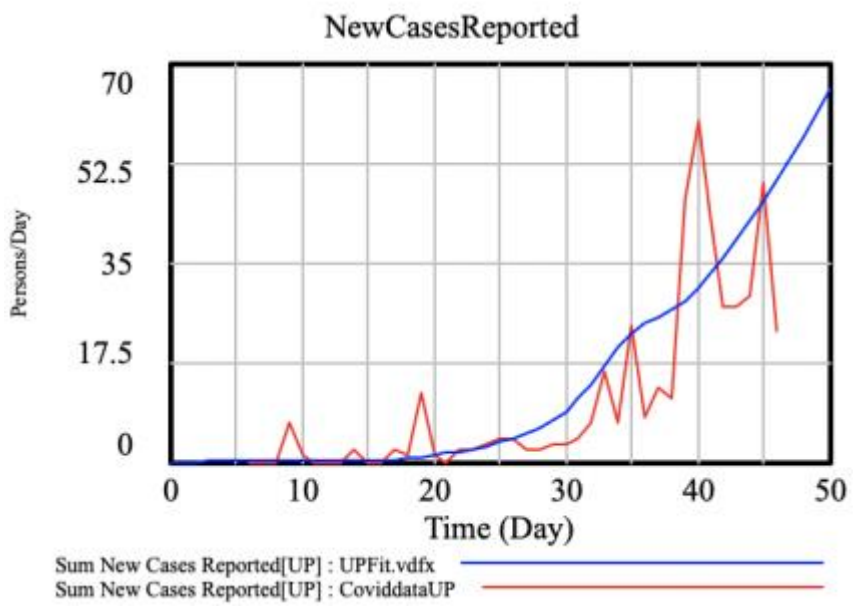
</td><td>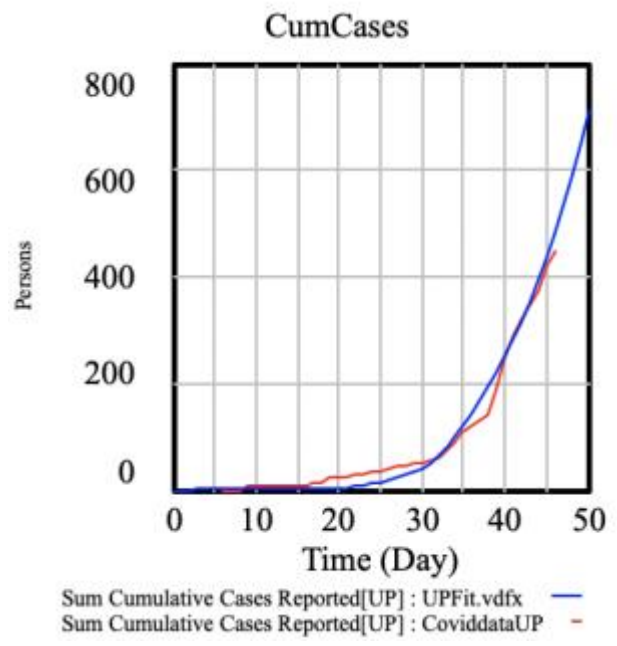
</td><td>$f_4(t) = 0.25$<br><br>$x(t) = 0.1$<br><br>$y(t) = \begin{cases} 1.9 & t \leq 15 \\ 1.9 - (\frac{0.7}{35}(t-15)) & 15 < t < 50 \end{cases}$</td></tr>
<tr><td>West Bengal</td><td></td><td></td></tr>
<tr><td>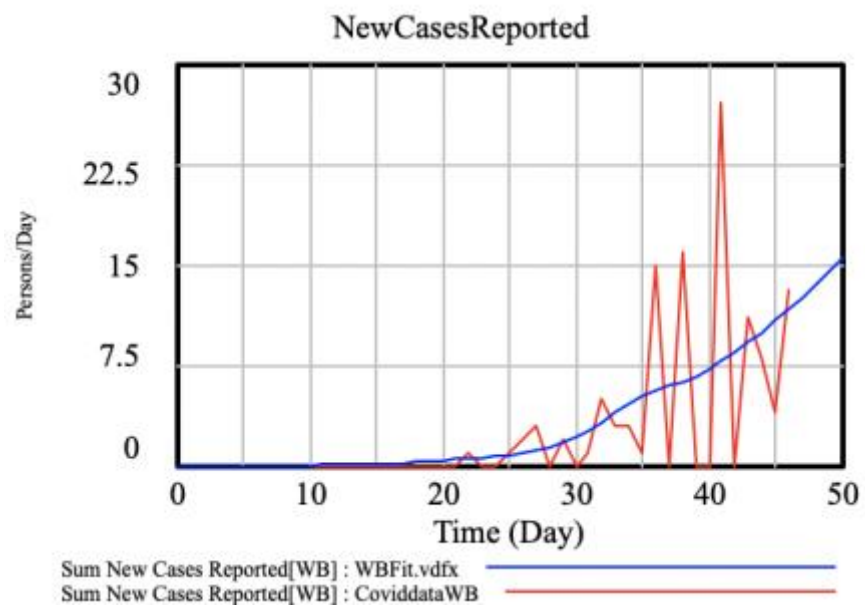
</td><td>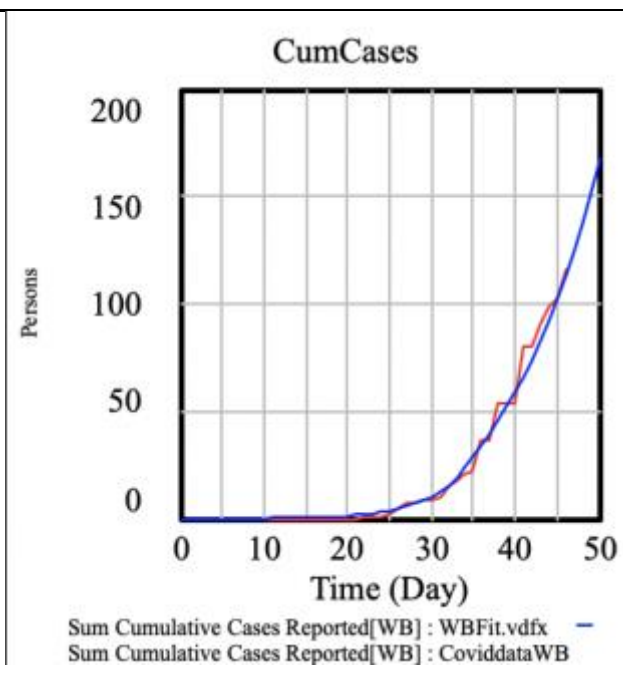
</td><td>$f_4(t) = 0.25$<br><br>$x(t) = 0.1$<br><br>$y(t) = \begin{cases} 2 & t \leq 15 \\ 2 - (\frac{0.8}{35}(t-15)) & 15 < t < 50 \end{cases}$</td></tr>
</table>